%% file: main.tex
\documentclass[journal ]{new-aiaa}
\usepackage{amsmath,amsfonts,subcaption}
\usepackage{algorithmic}
\usepackage{graphicx}
\usepackage{textcomp}
\usepackage{xcolor}
\usepackage{hyperref}
\usepackage{balance}

\usepackage[utf8]{inputenc}
\usepackage{textcomp}

\usepackage{graphicx}
\usepackage{amsmath}
\usepackage[version=4]{mhchem}
\usepackage{siunitx}
\usepackage{longtable,tabularx}

\newlength{\mintednumbersep}
\AtBeginDocument{%
  \sbox0{\tiny00}%
  \setlength\mintednumbersep{2\parindent}%
  \addtolength\mintednumbersep{-\wd0}%
}

\def\BibTeX{{\rm B\kern-.05em{\sc i\kern-.025em b}\kern-.08em
    T\kern-.1667em\lower.7ex\hbox{E}\kern-.125emX}}

\newtheorem{problem}{Problem}

\title{Early Design Exploration of 
Aerospace Systems Using Assume-Guarantee Contracts
}

\author{Nicolas Rouquette \footnote{Principal member of technical staff, Information Systems Division, AIAA Member}}
\affil{NASA Jet Propulsion Laboratory, California Institute of Technology, Pasadena, CA 91109, USA}
\author{Alessandro Pinto\footnote{Autonomy assurance lead, Office of Safety and Mission Success, AIAA Member }}
\affil{NASA Jet Propulsion Laboratory, California Institute of Technology, Pasadena, CA 91109, USA}
\author{Inigo Incer\footnote{Postdoctoral scholar research associate, Computing and Mathematical Sciences, AIAA Member}}
\affil{California Institute of Technology, Pasadena, CA  91125, USA}

\newcommand{\pDSN}[0]{\texttt{DSN}}
\newcommand{\pSBO}[0]{\texttt{SBO}}
\newcommand{\pTCM}[0]{\texttt{TCM}}
\newcommand{\pTCMh}[0]{\texttt{TCM\_h}}
\newcommand{\pTCMdv}[0]{\texttt{TCM\_dv}}
\newcommand{\pCHRG}[0]{\texttt{CHRG}}

\begin{document}

\maketitle

\input{abstract.tex}

\section{Introduction}
\input{introduction}

\section{Overview of Contracts \& Pacti}
\label{sc:pacti}
\input{pacti-overview}

\section{Small-body asteroid mission}
\label{sc:mission}
\input{mission}

\section{Preliminary design of an aircraft fuel system}
\label{sc:aircraft-example}
\input{aircraft}

\section{Concluding remarks}
\label{sc:conclusions}
\input{conclusions}

\bibliography{references}

\end{document}

%% file: abstract.tex
\begin{abstract}
We present a compositional approach to early modeling and analysis of complex aerospace systems based on assume-guarantee contracts. 
Components in a system are abstracted into assume-guarantee specifications. Performing algebraic contract operations with Pacti allows us to relate local component specifications to that of the system.
Applications to two aerospace case studies---the design of spacecraft to satisfy a rendezvous mission and the design of the thermal management system of a prototypical aircraft---show that this methodology provides engineers with an agile, early analysis and exploration process.
\end{abstract}

%% file: introduction.tex
In the early phases of a typical design process, engineers focus on defining an initial set of requirements and allocating them to subsystems and components. This requirement definition process is challenging due to the need for balancing goals and constraints from customers (top-down) against the capabilities of available or implementable components (bottom-up)\cite[Ch. 3]{NPR7123.1D}. Stringent requirements may be infeasible due to technological limitations or can lead to over-design\cite{SEPrinciplesPractice}, overly loose requirements risk producing a system that fails to meet customer goals or lacks robustness against changes in those goals\cite{DFC}.

Systems, sub-systems, and component engineers must negotiate specifications in this early phase. This process involves frequent interactions among subject-matter experts (SMEs) to elicit requirements, a notoriously challenging activity\cite{8049166}. Furthermore, the heterogeneous nature of this activity makes collaboration and negotiation difficult \cite{10.1109/WICT.2013.7113143}. Prototyping \cite{6828119, SEPrinciplesPractice, NPR7123.1D} and simulation \cite{SEPrinciplesPractice, NPR7123.1D} allow combining subsystem models into system-level models to explore various scenarios. However, several challenges arise with simulation-based analyses during the early design stages. Firstly, sufficiently detailed subsystem simulation models are rarely available when subsystem specifications are still evolving. Secondly, even if such models exist, few experts within the organization possess the skills to utilize them effectively. In particular, managing a system-level simulation that integrate subsystem-level models from all SMEs is beyond the expertise of any single individual. Thirdly, detailed simulations often require significant software setup and runtime, limiting their ability to provide insights into multiple what-if scenarios quickly. Lastly, the results obtained through simulation are typically valid only for specific system implementations under specific operating conditions. Thus, while insightful, analysis by simulation remains an incomplete and time-consuming analysis method poorly suited for  early design stage elicitation. 

To support the early stage of system design, we seek an alternative prototyping methodology with the following characteristics:
\begin{enumerate}
\item Implementation flexibility: The methodology and supporting tools should operate on a \emph{a range of possible implementations} for each component used in the design. 
By considering sets of components instead of a specific implementation, we obtain 
\emph{implementation flexibility}, ensuring that analysis results are valid for all valid implementations of the models representing each element. This conservative flexibility prevents engineers from being cornered into sub-optimal solutions.

\item Sound and Complete Analysis Procedures: Unlike simulation-based analyses, where a successful run does not guarantee desired characteristics for all possible executions, the methodology should assert that systems will exhibit the desired properties for all potential runs.

\item Support for Compositional Reasoning: Compositional reasoning allows the decomposition of analysis and verification problems into smaller, more manageable tasks, addressing the computationally prohibitive nature of analyzing complex systems. It also enables the independent implementation of subsystems and components. Once a set of requirements has been defined and allocated, and it has been demonstrated that these requirements satisfy top-level requirements, different teams or suppliers can independently implement these components. As long as each component satisfies its local requirements, the overall system is guaranteed to meet top-level goals. This formal compositional methodology is essential for handling the increasing number of requirements and avoiding integration problems arising from informal approaches.

\item Fast Turnaround and Insightful Feedback: In the early phases, the methodology should enable quick transitions from candidate designs to figures of merit and swift verification that the requirement breakdown meets system objectives. When system objectives are not met, the tools should provide explanations for the shortcomings. Automatic verification tools should offer insights into system properties, moving beyond binary feedback to compute the model of the entire system from component models. This approach allows systems engineers to understand why the system fails to meet top-level requirements, facilitating better-informed decision-making.

\end{enumerate}

These characteristics motivate us to explore the application of a formal and compositional framework in the early stages of design, focusing on \emph{formal requirements} as the specification of components. Requirements represent multiple implementations, satisfying (1) above. Using requirements means specifying the properties components must satisfy without being prescriptive about how they satisfy them. We adopt a modeling framework based on the theory of \emph{assume-guarantee contracts} \cite{BenvenisteContractBook}. Contracts are formal specifications split in two parts:
the component's guarantees and
the component's expectations of its operational context to deliver its guarantees.
Contracts have several algebraic operations that enable us to carry out compositional system-level analysis \cite{Incer:EECS-2022-99}, satisfying (3) and (4). The analysis enabled by these operations is both sound and complete, as specified in (2).

In this paper, we will represent the assumptions and guarantees of a contract using polyhedral constraints over the variables that describe the interface of a component. We use the Pacti~\cite{pacti} tool for modeling and automatic analysis since it meets the above-mentioned requirements. Pacti allows us to model systems using assume-guarantee contracts and can compute explicit representations of several of contracts\footnote{Pacti is available as an open-source package under a BSD 3-Clause license at \url{https://github.com/pacti-org/pacti}}. Pacti is designed with computational efficiency and explainability as top priorities. Consequently, a designer can use it interactively to gather quick feedback (less than three seconds in our experiments) about the impact of varying design parameters. Thus, we believe the system designer can benefit significantly by specifying subsystem operational behaviors as contracts and manipulating them using Pacti.

The purpose of this article is to show that the use of contracts, their algebraic operations, and their tool support leads to the effective exploration of design alternatives for aerospace systems. We accomplish this by applying these techniques to early design exploration of two case studies: $(i)$ a CubeSat-sized spacecraft performing a small-body asteroid rendezvous mission, and $(ii)$ the fuel and thermal management system of a hypothetical aircraft. For the former, we model mission scenarios as sequences of task-specific steps, showing how each step can be modeled by a contract and demonstrating how the composition and merging operations are used to compute mission-level metrics as a function of different spacecraft parameters. We use the notion of viewpoints to split a complex contract into simple subcontracts, each focused on one aspect of the task, such as power, science \& communication, and navigation. Once we have defined the contracts for each task, we use Pacti to perform the following tasks: sequencing task-specific steps using contract composition, fusing such sequences across viewpoints using contract merging, computing figures of merit such as bounds on the average battery state of charge across a sequence using optimization, and evaluating bounds for state variables at arbitrary steps in the sequence. 

For the fuel and thermal management systems, we model the key components as parametric contracts. The parameters represent tolerance values of component properties, correlating with the quality of the implementations of these components. The system includes other parameters that capture the operating point, including internal variables such as fuel flow rates and external variables such as flight altitude and required level of thrust. We use Pacti to compute the composition of the component contracts and the expected temperature bounds in key points of the thermal management system. This exploration is used to select promising operating points for further optimization. We integrate Pacti with an optimization procedure to find the largest acceptable tolerances for the components that still guarantee the satisfaction of safety margins.

\textbf{Related work.}
\input{related-work}

\textbf{Paper outline.} Section~\ref{sc:pacti} provides an overview of contracts and Pacti. Section~\ref{sc:mission} presents a case study of early design exploration of a space mission operation involving a small-body asteroid. We model subsystem-specific tasks as contracts and use Pacti to obtain insight into many system-level aspects. Section~\ref{sc:aircraft-example} presents the use of contracts and Pacti in the analysis and design of the fuel and thermal management system of an aircraft.
We conclude in Section~\ref{sc:conclusions}.

%% file: related-work.tex
Several computer-based frameworks for contract-based modeling and verification, such as AGREE \cite{AGREE} and OCRA \cite{OCRA}, have been developed over the past decade. In AGREE, assumptions and guarantees are specified in a synchronous language, while in OCRA they are specified in Linear Temporal Logic. These systems allow users to instantiate and connect components and to check whether such composition refines a higher-level contract using a series of satisfiability queries~\cite{cimatti2012property}. In contrast, Pacti~\cite{pacti} explicitly supports multiple viewpoints and the ability to compute the result of the composition operation in the form of a new contract at the interface of the system. This capability provides the actual expression of the contract implemented by the composition of components, resulting in two advantages: (1) it allows users to inspect the result of the composition and to gain insights into the reasons why a system satisfies (or not) its specifications, and (2) it allows to use the result of the composition to compute the space of accepted inputs (or the space of possible outputs) of the entire system, which can be used to design margins or compute robustness metrics. Moreover, Pacti implements algorithms that can compute the quotient of a specification with respect to a partial system to obtain the specification of a missing component.

We have used these features in the two aforementioned case studies. In the first case study, we used contracts to model actions and state transitions in a sequence of spacecraft operations. Several languages have been defined in the past to deal with sequences of operations. The AI planning community uses the Planning Domain Definition Language (PDDL) \cite{fox2003pddl2}, including extensions for hierarchical planning \cite{GEORGIEVSKI2015124}. These languages focus on succinctness and efficiency for the purpose of planning, and there is no tool support for composition and other algebraic operations.  On the other hand, the contract framework in Pacti~\cite{pacti} focuses on system and sub-system modeling from different viewpoints, and it is expressive enough to encode state transitions. The contract framework focuses on concurrent, component-based engineering of complex systems.  Planning/scheduling systems designed for embedded systems impose significant restrictions on the expressiveness of the planning language to ensure responsiveness in limited computing environments. In \cite{chien2012generalized,rabideau2017managing}, the authors describe a generalized timeline representation restricting  planning constraints to a single variable linear constraint. In contrast, Pacti supports linear constraints with multiple variables as shown in the case study above. 
Finally, the analysis presented in the first case study differs from classical planning and scheduling. In planning, we are given a set of task models, an initial state, and a goal state; the problem is to find a sequence of task instances that bring a system from the initial state to the goal state \cite{ghallab2004automated}. Developing the task models (also referred to as domain authoring) is a major contributor to efficiency, and verifying and validating such models are hard and time-consuming activities \cite{mccluskey2017engineering}. The analysis we present can be seen as addressing the joint exploration of system requirements allocation and domain authoring to derive optimal task specifications as a precursor to the development of task models.

Instead of using contracts for operational analysis, the second case study shows using contracts to design and analyze an aircraft sub-system. Some previous work in this area include~\cite{pinto2017csl4p} and ~\cite{6690099}. We use the same case study presented in ~\cite{pinto2017csl4p}. However, the analysis and design space exploration in this paper extends considerably that of  ~\cite{pinto2017csl4p}, whose focus is refinement checking, not design space exploration. This is an important difference because design space exploration requires the explicit computation of the performance of a composed system. The analysis presented in~\cite{6690099}, while spanning several abstraction layers, seems to focus on the use of ``vertical'' contracts, meaning on the propagation of assumptions and guarantees between one stage of a design process and the next. Instead, we focus on the analysis of requirement decomposition at one level, but by computing the result of the composition explicitly.

\textbf{Contributions.} To the best of our knowledge, this work, which is a continuation of our conference publication \cite{smcit23}, is the first to leverage the explicit solution of algebraic operations on contracts in the early analysis and design of aerospace systems. The use of compositional methods and specifically contract-based design has been mainly centered around addressing the refinement verification problem. Namely, given a system specification $S$ as a pair of assumptions $A$ (the set of environments of interest) and guarantees $G$ (the valid set of behavior of an implementation for an environment in $A$), a designer architects a system of interconnected components $\{C_1,\ldots,C_n\}$, and checks whether $S$ is satisfied by the composition without having to compute it. 
In contrast, this work exploits the explicit computation of contract operations using Pacti in the design space exploration of aerospace systems. In addition to the space mission case study described in \cite{smcit23}, this paper includes a case study on the fuel and thermal management system of a hypothetical aircraft.

%% file: pacti-overview.tex
Pacti \cite{pacti} helps designers to reason about specifications and to manipulate them. These specifications are given to Pacti as assume-guarantee contracts, which are pairs $(A,G)$ where $A$ is a set of assumptions, and $G$ a set of guarantees. Contracts resemble the form in which datasheets are typically written, i.e., a component's datasheet specifies that the component will satisfy certain guarantees only when its context of operation satisfies certain assumptions. This section provides a brief overview about Pacti.

For Pacti, a contract has four elements:
\begin{itemize}
    \item A set of \emph{input variables}.
    \item A set of \emph{output variables}.
    \item A set of \emph{assumptions} that are constraints on the input variables.
    \item A set of \emph{guarantees} that are constraints on both input and output variables. 
\end{itemize}

Intuitively, the assumptions define a set of possible environments in which the subsystem can be used. The guarantees define the input-output relation that the subsystem promises to enforce when the assumptions are satisfied. Pacti currently supports constraints expressed as linear inequalities, also called \emph{polyhedral constraints}, but its architecture is extensible to other constraint formalisms.

The algebra of contracts has been formalized in several previous works (see, for instance~\cite{BenvenisteContractBook,Incer:EECS-2022-99}, and references therein). The formalization includes the definition of several operators and their properties. These operators can be used to address several tasks relevant to system design, including: 

\begin{itemize}
    \item 
Building systems out of subsystems. Suppose that we have specified contracts for a set of subsystems. We can define a system as the assembly of such subsystems. The operation of \emph{composition} allows us to compute the contract of such a system from the contracts of the assembled subsystems. In other words, the composition operator provides a mechanism for computing system contracts from subsystem contracts.
\item 
Patching systems. The operation of \emph{quotient} allows us to compute the contract of a subsystem that needs to be composed with an existing subsystem so that the resulting system composition meets a top-level contract. In other words, the quotient finds contracts of missing subsystems from contracts for the system and a partial implementation.
\item
Validating decompositions. \emph{Refinement} allows us to tell when a contract is more relaxed, or less demanding than another. When a subsystem satisfies a contract, it is guaranteed to satisfy a more relaxed contract. When a system contract is broken into an assembly of subsystem contracts, refinement allows us to tell whether this decomposition is a valid refinement of the system-level contract.
\item
Fusing viewpoints. The operation of \emph{merging} allows us to generate a single contract whose assumptions and guarantees require the satisfaction of the assumptions and guarantees of the merged contracts, respectively. In other words, merging fuses multiple contract viewpoints, a common operation in concurrent design.
\end{itemize}

The following sub-sections provide an overview of how Pacti supports these common tasks.

\begin{figure}
\centering
\begin{subfigure}[t]{0.40\columnwidth}
\centering
\includegraphics[width=\textwidth]{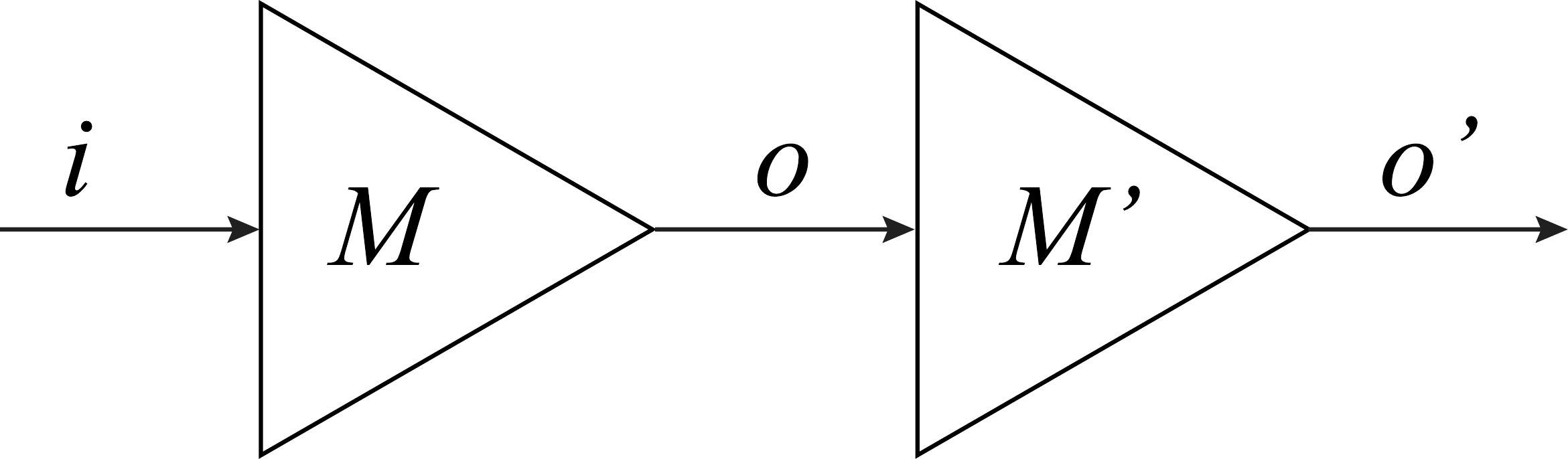}
\end{subfigure}
\quad
\begin{subfigure}[t]{0.40\columnwidth}
\centering
\includegraphics[width=\textwidth]{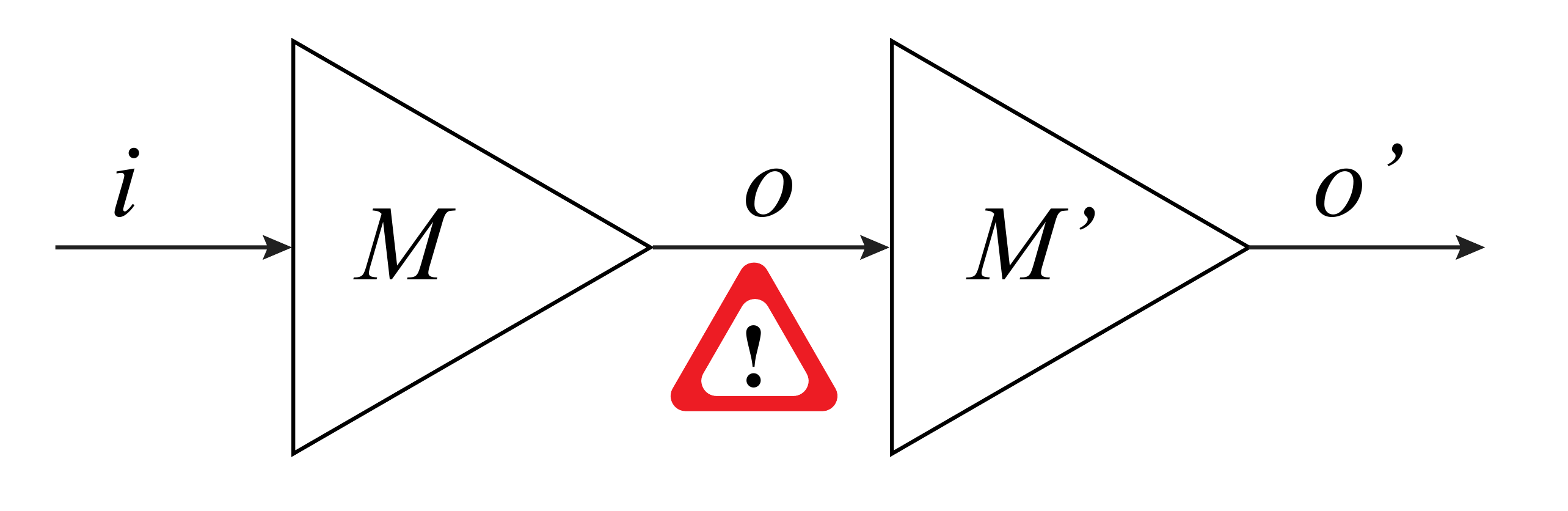}
\end{subfigure}

\begin{subfigure}[t]{0.40\columnwidth}
\centering
\begin{verbatim}
c1 =
    PolyhedralContract.from_string(
    InputVars=["i"],
    OutputVars=["o"],
    assumptions=["|i| <= 2"],
    guarantees=["o - i <= 0",
                "i - 2o <= 2"])

c2 =
    PolyhedralContract.from_string(
    InputVars=["o"],
    OutputVars=["o_p"],
    assumptions=["o <= 0.2",
                 "-o <= 1"],
    guarantees=["o_p - o <= 0"])

sys_contract = c1.compose(c2)
print(sys_contract)
\end{verbatim}

\end{subfigure}
\quad
\begin{subfigure}[t]{0.40\columnwidth}
\centering
\begin{verbatim}

c1_n =
PolyhedralContract.from_string(
    InputVars=["i"],
    OutputVars=["o"],
    assumptions=["|i| <= 2"],
    guarantees=["|o| <= 3"])

c2 =
    PolyhedralContract.from_string(
    InputVars=["o"],
    OutputVars=["o_p"],
    assumptions=["o <= 0.2",
                 "-o <= 1"],
    guarantees=["o_p - o <= 0"])

new_sys_contract = c1_n.compose(c2)
\end{verbatim}
\end{subfigure}

\vspace{10px}

\begin{subfigure}[t]{0.40\columnwidth}
\centering
\begin{verbatim}
# Output:
InVars: [i]
OutVars:[o_p]
A: [
    i <= 0.19999999999999996
    -0.5 i <= 0.0
]
G: [
    -i + o_p <= 0.0
]
\end{verbatim}
\end{subfigure}
\quad
\begin{subfigure}[t]{0.40\columnwidth}
\centering
\begin{verbatim}
# Output:
IncompatibleArgsError: Could not
eliminate variables ['o'] by
refining the assumptions 
[
    o <= 0.19999999999999996
    -o <= 1.0
]
using guarantees
[
    |o| <= 3.0
]
\end{verbatim}
\end{subfigure}

\begin{subfigure}[b]{0.40\columnwidth}
\centering
\caption{}
\label{fg:composition}
\end{subfigure}
\begin{subfigure}[b]{0.40\columnwidth}
\centering
\caption{}
\label{fg:diagnostics}
\end{subfigure}

\caption{(a) System composition (b) System diagnostics}
\end{figure}

\begin{figure}
\centering
\begin{subfigure}[t]{0.40\columnwidth}
\centering
\includegraphics[width=\textwidth]{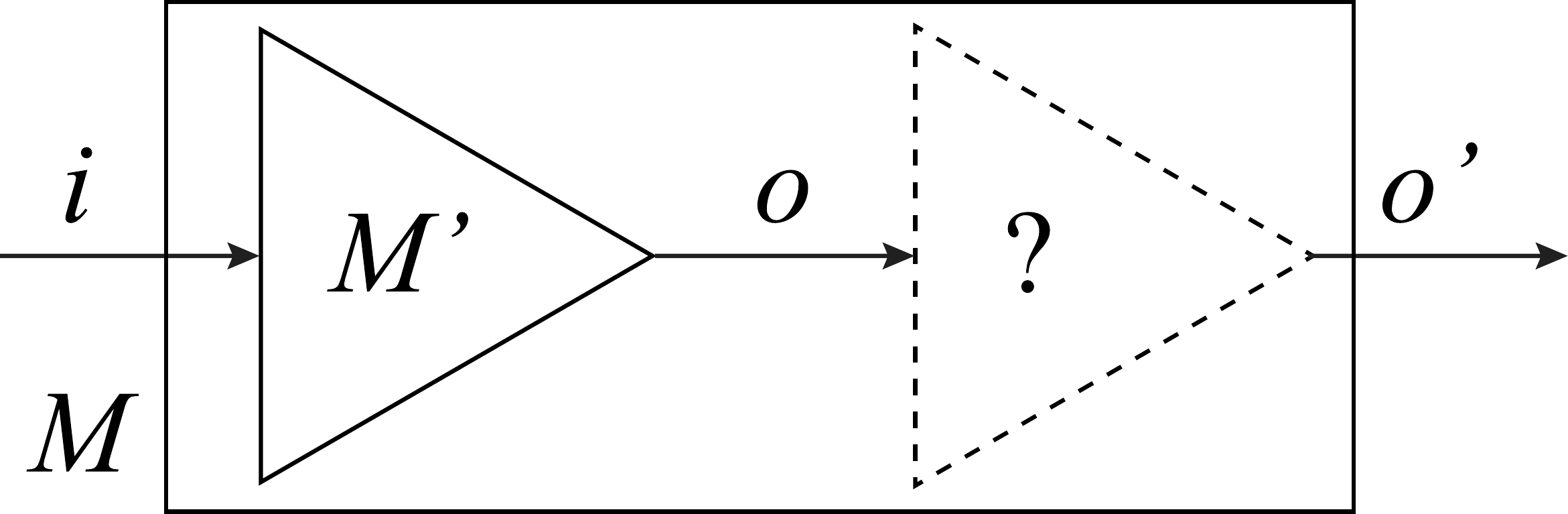}
\end{subfigure}
\begin{subfigure}[t]{0.40\columnwidth}
\centering

\includegraphics[width=0.8\textwidth]{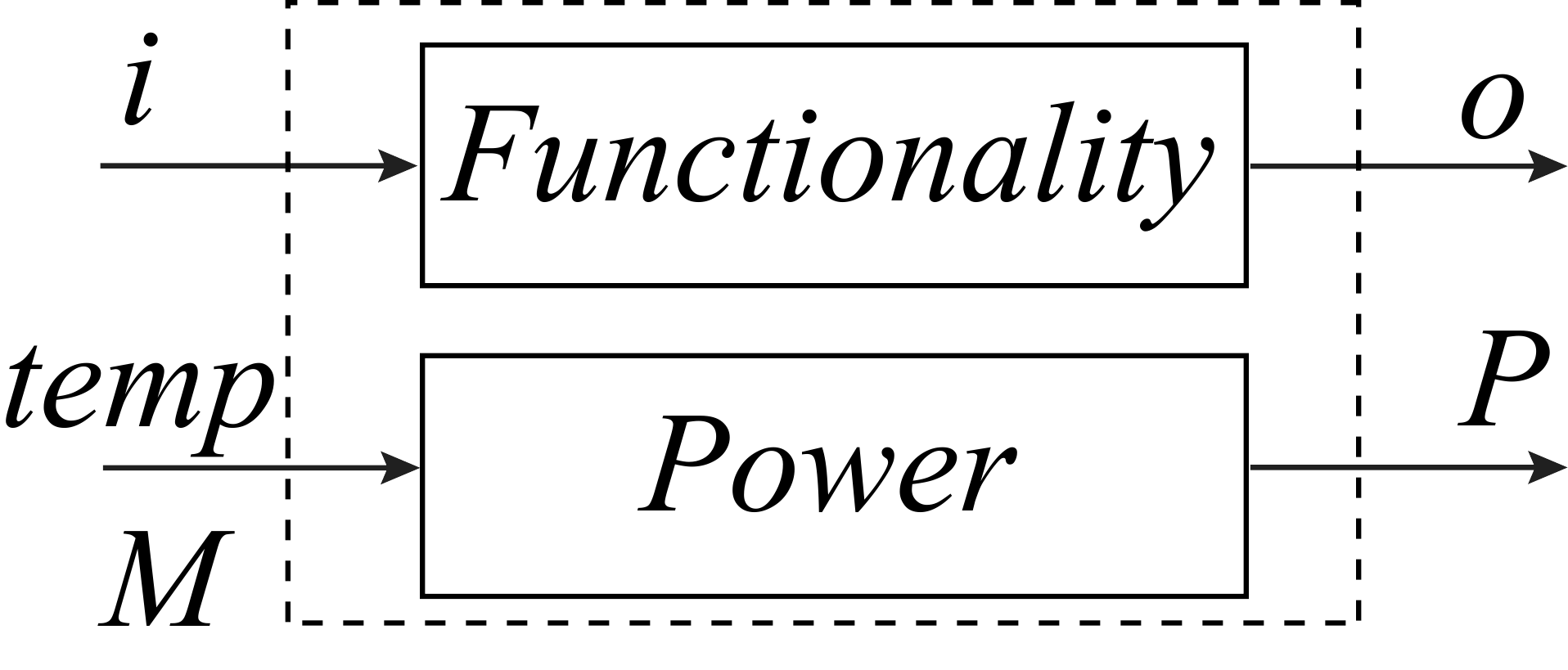}
\end{subfigure}
\begin{subfigure}[t]{0.40\columnwidth}
\centering
\begin{verbatim}
c_top =
    PolyhedralContract.from_string(
    InputVars=["i"],
    OutputVars=["o_p"],
    assumptions=["|i| <= 1"],
    guarantees=["o_p - 2i = 1"])

c_partial =
    PolyhedralContract.from_string(
    InputVars=["i"],
    OutputVars=["o"],
    assumptions=["|i| <= 2"],
    guarantees=["o - 2i = 0"])

c_missing = c_top.quotient(c_partial)
print(c_missing)
\end{verbatim}
\end{subfigure}
\quad
\begin{subfigure}[t]{0.40\columnwidth}
\centering
\begin{verbatim}
funct_vp =
    PolyhedralContract.from_string(
    InputVars=["i"],
    OutputVars=["o"],
    assumptions=["|i| <= 2"],
    guarantees=["o - 2i = 1"])

pwr_vp =
    PolyhedralContract.from_string(
    InputVars=["temp"],
    OutputVars=["P"],
    assumptions=["temp <= 90"],
    guarantees=["P <= 2.1"])

sys_cont = funct_vp.merge(pwr_vp)
print(sys_cont)
\end{verbatim}
\end{subfigure}

\vspace{10px}

\begin{subfigure}[t]{0.40\columnwidth}
\centering
\begin{verbatim}
# Output:
InVars: [o]
OutVars:[o_p]
A: [
    |o| <= 2.0
]
G: [
    -o + o_p = 1.0
]
\end{verbatim}
\end{subfigure}
\quad
\begin{subfigure}[t]{0.40\columnwidth}
\centering
\begin{verbatim}
# Output:
InVars: [i, temp]
OutVars:[o, P]
A: [
  |i| <= 2.0
  temp <= 90.0
]
G: [
  -2.0 i + o = 1.0
  P <= 2.1
]
\end{verbatim}
\end{subfigure}

\begin{subfigure}[b]{0.40\columnwidth}
\centering
\caption{}
\label{fg:patch}
\end{subfigure}
\begin{subfigure}[b]{0.40\columnwidth}
\centering
\caption{}
\label{fg:merging}
\end{subfigure}

\caption{(a) Patching systems (b) Fusing viewpoints}
\end{figure}

\subsection{Computing system specifications}
\label{sec:composing-specifications}

Consider the system shown in Figure~\ref{fg:composition}. Subsystem $M$ has input $i$ and output $o$, and $M'$ has input $o$ and output $o'$. The assumptions and guarantees of $M$ are, respectively, $\{|i| \le 2\}$ and $\{o \le i \le 2o + 2\}$, while the assumptions and guarantees of $M'$ are, respectively, $\{-1 \le o \le 0.2\}$ and $\{o' \le o\}$.
The figure shows how we can use Pacti to obtain the contract of the system formed by assembling these two subsystems. Pacti tells us that the system contract has input $i$, output $o'$, assumptions $\{0 \le i \le 0.2\}$, and guarantees $\{o' \le i\}$. Pacti's answer only involves the top-level input and output variables, having eliminated the intermediate variable, $o$.

\subsection{System diagnostics}
\label{knqfqrwbg}

In Figure~\ref{fg:diagnostics}, we have the same subsystems as those shown in Figure~\ref{fg:composition}, except that the guarantees of $M$ have been replaced by $\{|o| \le 3\}$. When we try to form a system using $M$ and $M'$, Pacti tells us that the guarantees of $M$ are insufficient to satisfy the assumptions of $M'$. Indeed, $M'$ requires its input to be bounded by 0.2, while $M$ guarantees this signal to be bounded by 3. Pacti thus flags a potential flaw in our design.

\subsection{Specifying missing subsystems}

Figure~\ref{fg:patch} shows the situation in which we will implement a system $M$ with input $i$ and output $o'$ having assumptions $\{|i| \le 1\}$ and guarantees $\{o' = 2i + 1\}$.
To implement this system, we will use a subsystem $M'$ with input $i$, output $o$, assumptions $\{|i| \le 2\}$ and guarantees $\{o = 2i\}$.
To implement the top-level specification using $M'$, we have to identify the specification of the missing subsystem denoted by a question mark in the figure. Pacti computes this missing-subsystem specification for us, saying that this subsystem will have input $o$, output $o'$, assumptions $\{|o| \le 2\}$ and guarantees $\{o' = o + 1\}$.

\subsection{Fusing viewpoints}
\label{sec:fusing-viewpoints}
Contract-based design enables us to organize specifications in categories, or viewpoints.
Figure~\ref{fg:merging} show a subsystem $M$ with two different contracts assigned to it: a functionality contract and a power contract.
The operation of merging can generate a single contract that contains both viewpoints of the design.
When performing analysis, we use only the subsystem specifications for the task at hand.
For example, to carry out power analysis of an entire system, we should be able to use only the power viewpoints of the subsystems that compose it.

%% file: mission.tex
\begin{figure}
\centering
\includegraphics[width=0.8\textwidth]{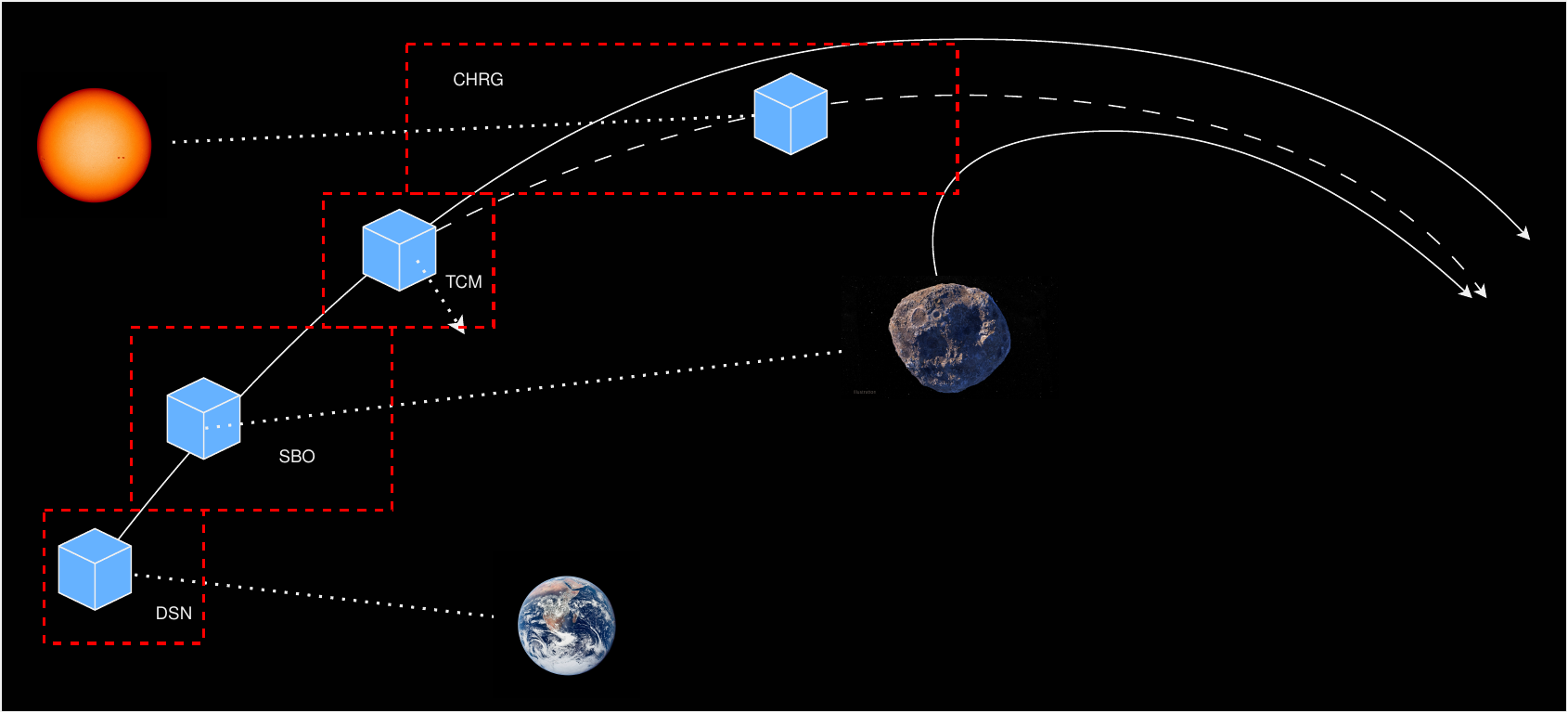}
\caption{Segmented space mission scenario for asteroid approach}
\label{fig:segmented-scenario}
\end{figure}

In this section, 
we focus on the problem of operating a CubeSat-sized spacecraft performing a small-body asteroid rendezvous mission\footnote{This case study is available at \url{https://github.com/pacti-org/cs-space-mission}, tag: SMC-IT-2023.}. Figure \ref{fig:segmented-scenario} illustrates a simplification of the small-body asteroid approach scenario described in more detail in \cite{nesnas2021autonomous}\footnote{The Sun, Earth, spacecraft, and small-body asteroid are shown at different scales for illustration purposes.}. During its mission, the spacecraft (blue cube) has the high level objective of approaching an asteroid, making measurements of scientific interest, and sending this data to earth.
To achieve the mission, the spacecraft must perform a sequence of the following basic tasks.
To communicate with Earth, the spacecraft must orient its fixed antennas towards Earth (task \pDSN). Depending on the trajectory, this orientation may be suboptimal for the spacecraft panels to produce maximum electrical power. When energy is needed, the spacecraft must find a way to reorient itself towards the sun (task \pCHRG). Optical science measurements also require orienting the spacecraft's camera towards the asteroid for observation (task \pSBO). Finally, autonomous navigation requires a different orientation to yield the desired velocity change when performing a Trajectory Correction Maneuver (task \pTCM).
Each of these tasks will be characterized according to certain parameters $P$. Among these parameters, we will have 
energy consumption rates, energy generation rates, rate of convergence for trajectory correction maneuvers, etc.

\begin{figure}
\centering
\includegraphics[width=0.8\textwidth]{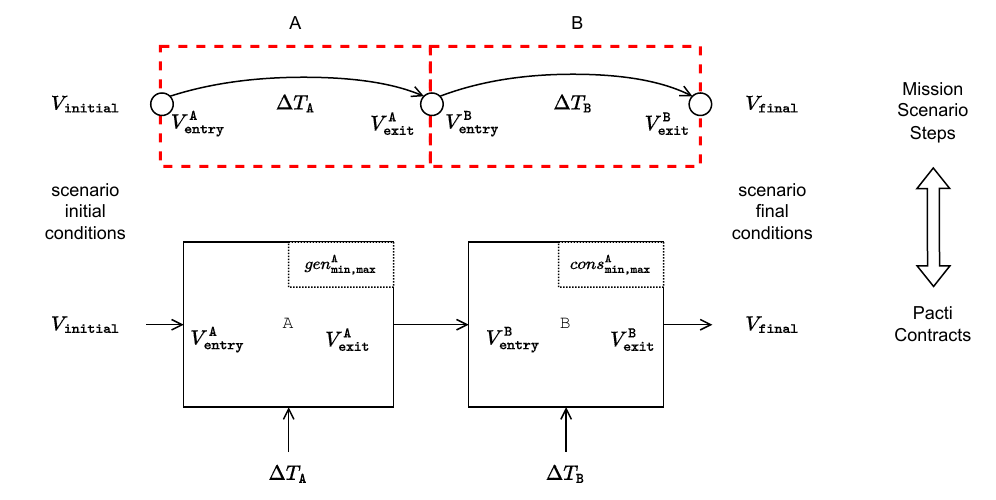}
\caption{Contract modeling of scenario steps and sequences}
\label{fig:2steps}
\end{figure}

\begin{problem}[Mission analysis and design]
\label{pr:mission-analysis}
We will be interested in characterizing
sets of task parameters that will ensure that the spacecraft is able to complete mission requirements. The mission objectives will require the spacecraft to record and transmit to earth a certain amount of scientific data, and to operate with its battery level never falling below a certain threshold.
\end{problem}

\subsection{Leveraging contracts for mission analysis and design}

Over the course of the operation of the spacecraft, we will be interested in tracking the values of the following quantities of interest, or state variables. These state variables will be used to state mission-level requirements.
\begin{itemize}
    \item The state of charge of the battery, denoted by $soc$. Its value will be a percentage.
    \item The amount of onboard science data storage, denoted by $d$. This value will be a percentage. 100\% will mean that all onboard storage is used to hold the measurements that have been gathered and not yet transmitted to earth.
    \item Cumulative science data acquired, denoted by $c$. This is positive real number. It represents the total amount of data gathered over the course of the mission.
    \item Relative trajectory estimation uncertainty, denoted by $u$ and given as a percentage.
    \item Relative trajectory progress, denoted by $r$ and given as a percentage.
\end{itemize}

As shown in previous work on planning and scheduling for space missions~\cite{chien2012generalized,rabideau2017managing,chen2003automated}, the effect of high-level tasks on the spacecraft's states can often be represented by linear inequalities of the form $a\cdot t \leq b$ where $t$ is a time, $a$ is a rate constant, and $b$ is a value constant. State variables that are typically represented using linear constraints include the state of charge and the data generated by science experiments or sent to Earth. This class of constraint formulas fits within the expressiveness of Pacti's polyhedral constraints. Thus, we explore modeling tasks as assume-guarantee contracts in Pacti. 

Figure \ref{fig:2steps} illustrates how tasks are modeled as contracts, and how sequences are derived through the composition~\ref{sec:composing-specifications} of task models. Each task instance such as \texttt{A} or \texttt{B} in Figure~\ref{fig:2steps} is a contract specifying the system behavior during that step of a mission scenario. The contract defines the entry and exit conditions on the state variables of the spacecraft as assumption and guarantee constraints, respectively, for a given duration $\Delta T$ of the mission step.
For example, task \texttt{A} specifies assumptions in terms of its input variables $\{V_{\texttt{entry}}^\texttt{A}, \Delta T_\texttt{A}\}$ only, and guarantees in terms of both input and output variables $\{V_{\texttt{entry}}^\texttt{A}, V_{\texttt{exit}}^\texttt{A}, \Delta T_\texttt{A}\}$. Time elapses while moving from left to right in Figure~\ref{fig:2steps}: the inputs of a contract represent the state at time $t$, while the outputs represent the state at time $t + \Delta T_\texttt{A}$. Sequences are obtained by linking the output variables of a contract representing a mission step to the input variables of the next step.

Schedulability analysis (see Problem~\ref{pr:mission-analysis}) is the exploration of sequences that together refine system-level specification. The requirements specify constraints on the initial conditions, final conditions, duration, and performance at different points in the mission. We leverage the capability of Pacti to compute the composition of contracts and gain insights on the achieved mission level performance. We then implement a search of the hyperparameters $P$ of the model to define the requirements for each task. The hyperparameters of a generic task \texttt{A} include minimum and maximum power generation $gen_{\texttt{min,max}}^\texttt{A}$ for \texttt{A}, and minimum and maximum power consumption $cons_{\texttt{min,max}}^\texttt{B}$ for \texttt{B}. Sampling techniques can be used to generate multiple scenarios and to perform schedulability analysis. Our experience so far is that the computational complexity of this approach remains within practical considerations given that analysis in Pacti requires solving a number of linear programming problems proportional to the number of constraints involved. This means that constructing the contract for a scenario involving a finite number of steps to be composed and a finite number of operational requirements to be refined will result in a fixed upper bound on the number of linear programming problems to be solved. Since computing schedulability analyses over hyperparameter samples is easily parallelizable, this exploratory methodology enables a rapid turnaround between scenario contract modeling and analysis results.

\input{scenario-tasks}

\input{schedule-analysis}

%% file: scenario-tasks.tex
\subsection{Modeling the space mission using polyhedral assume-guarantee contracts.}
\label{sec:scenario-tasks}

\newcommand{\pgen}[0]{\text{pgen}}
\newcommand{\psgen}[0]{\text{sgen}}
\newcommand{\psoc}[0]{\text{soc}}
\newcommand{\pcons}[0]{\text{cons}}
\newcommand{\prate}[0]{\text{rate}}
\newcommand{\pnoise}[0]{\text{noise}}
\newcommand{\pimp}[0]{\text{imp}}

\newcommand{\pentry}[0]{\texttt{entry}}
\newcommand{\pexit}[0]{\texttt{exit}}
\newcommand{\pmin}[0]{\texttt{min}}
\newcommand{\pmax}[0]{\texttt{max}}
\newcommand{\pminmax}[0]{\texttt{min,max}}

\newcommand{\pCST}[0]{\texttt{CST}}
\newcommand{\pIn}[0]{\texttt{In}}
\newcommand{\pOut}[0]{\texttt{Out}}
\renewcommand{\pA}[0]{\texttt{A}}
\newcommand{\pG}[0]{\texttt{G}}

We will assume that, at any given time over the course of the mission, the spacecraft is executing one out of the following four tasks shown in Figure \ref{fig:segmented-scenario}:
\begin{description}
\item[\pDSN] Orient the spacecraft's antenna towards Earth to downlink science data.
\item[\pSBO] Orient the spacecraft's camera towards the asteroid for science and navigation observations.
\item[\pTCM] Orient the spacecraft's chemical thrusters in a direction to perform a Trajectory Correction Maneuver computed onboard to bring the spacecraft's trajectory closer to that of the asteroid.
\item[\pCHRG] Orient the spacecraft's solar panels towards the Sun to charge the battery.
\end{description}

\begin{table}
    \centering
    \begin{tabular}{|c|c|c|c|c|c|}
        \hline
         \textbf{Viewpoint} & \textbf{State} & \textbf{DSN} & \textbf{SBO} & \textbf{TCM} & \textbf{CHRG} \\
         \hline
         Power & $soc$ & $-$ & $-$ & $-$ & $+$ \\
         \hline
         Science \& & $d$ & $-$ & $+$ & $0$ & $0$ \\
         \cline{2-6}
         Communication & $c$ & $0$ & $+$ & $0$ & $0$ \\
         \hline
         Navigation & $u$ & $+$ & $-$ & $+$ & $+$ \\
         \cline{2-6}
         & $r$ & $0$ & $0$ & $+$ & $0$ \\
         \hline
    \end{tabular}
    \caption{Qualitative Task Impacts}
    \label{tab:qualitative_impacts}
\end{table}

Table \ref{tab:qualitative_impacts} summarizes the qualitative impacts of each type of task on key mission parameters, where $+$, $0$, and $-$ denote, respectively, positive, independent, and negative correlation of state variables with respect to task duration. In this table, we also group state variables by \emph{viewpoint}. Viewpoints are aspects of concern in the design process, such as power, timing, etc. In this case study, we will write contracts for each viewpoint of each task. 

To describe the Pacti contracts we authored in the case study, we name constants to convey their nature and adopt the following concise notation:

\begin{itemize}
    \item $x \in \begin{aligned}[t]& [\gamma_{\pmin\cdots\pmax}]\Delta T =  
    \{x \;|\; \gamma_{\pmin} \Delta T \le x \le \gamma_{\pmax} \Delta T\}\end{aligned}$
    \item $[\gamma_{\pmin\cdots\pmax}] = [\gamma_{\pmin}, \gamma_{\pmax}]$
    \item $v_{\pexit-\pentry} = v_{\pexit} - v_{\pentry}$
\end{itemize}

In this notation, $x$ is an expression; $\gamma_{min}$ and $\gamma_{max}$ are constants; and $v_{entry}$ and $v_{exist}$ are variables. All tasks assume valid ranges of input variables: task duration must be positive, if applicable, and other inputs must be within valid ranges. 

\begin{figure*}
\centering
\includegraphics[width=0.8\textwidth]{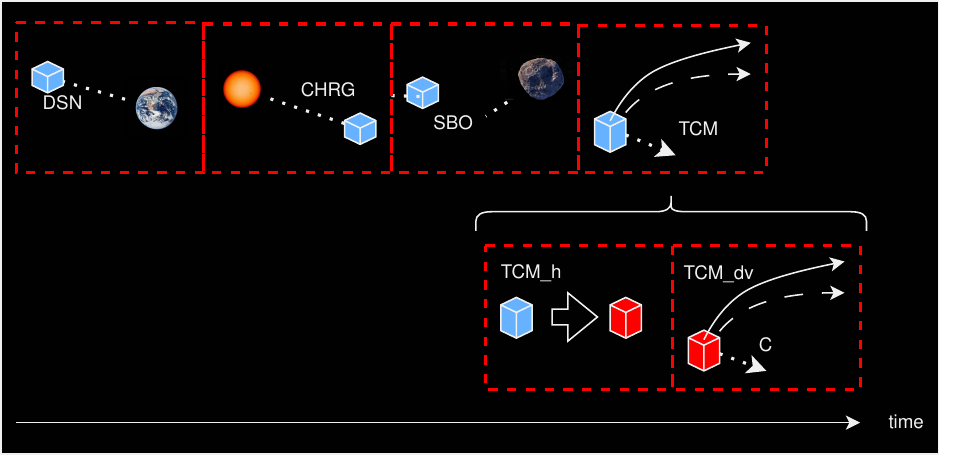}
\caption{Composite contract modeling a 5-step scenario sequence}
\label{fig:5steps}
\end{figure*}

Before we can analyze the schedulability of a mission operation scenario against operational requirements in Section \ref{mission:schedule-analysis}, we need to construct the scenario contract. Figure \ref{fig:5steps} shows a representative operation scenario involving a sequence of the following tasks: \pDSN, \pCHRG, \pSBO, and \pTCM. We decompose \pTCM into two subtasks: heating, \pTCMh, and a delta-v maneuver, \pTCMdv. We leverage Pacti's support for fusing viewpoints and break down the specification of each task across the three viewpoints described above: power, science \& communication, and navigation.
Our task now is to define contracts for each task and for each of these viewpoints.

\subsubsection{Pacti contracts in the power viewpoint}
Qualitatively, \pDSN, \pSBO, and \pTCM~have similar power-consuming contracts, whereas \pCHRG~has a power-generation contract. The guarantees assert that the change in state of charge will be proportional to a generation or consumption rate applied for the task's duration. \pTCM~involves two different power-consuming behaviors, thruster heating and delta-V, modeled as two subcontracts: \pTCMh~and \pTCMdv. Thus, the 4-step scenario becomes the composition of 5 steps. 

The task template notation below uses the following abbreviations: \pCST~for constant hyperparameters, \texttt{In,Out} for input and output variables, respectively, and \texttt{A,G} for contract assumptions and guarantees, respectively.

\begin{center}
\begin{tabular}{|cc|}
\multicolumn{2}{c}{Task template for $\mathcal{T}$\texttt{=CHRG}}\\
\hline
\pCST & $\pgen_{\pminmax}^\mathcal{T}$ \\
\hline
\pIn & $\psoc_{\pentry}, \Delta T^\mathcal{T}$ \\
\hline
\pOut & $\psoc_{\pexit}$ \\
\hline
\pA & $\Delta T^\mathcal{T} \ge 0 \And \psoc_{\pentry} \ge 0$ \\
\hline
\pG & 
{$\begin{aligned}
\psoc_{\pexit -\pentry} & \in [\pgen_{\pmin\cdots\pmax}]\Delta T^\mathcal{T}\\
\psoc_{\pexit} & \in [0,100]
\end{aligned}$} \\
\hline
\end{tabular}
\end{center}

The constant hyperparameter, $\pgen_{\pminmax}^\mathcal{T}$, defines the range of power generation charging the battery during an instance of this task, which guarantees that the difference between the exit and entry state of charge will be proportional to the power generation rate interval constant, $[\pgen_{\pmin\cdots\pmax}^\mathcal{T}]$, times the task duration input variable, $\Delta T^\mathcal{T}$.

\begin{center}
\begin{tabular}{|cc|}
\multicolumn{2}{c}{Task templates for $\mathcal{T} \in \{\texttt{DSN,SBO,TCM\_h,TCM\_dv}\}$}\\
\hline
\pCST & $\pcons_{\pminmax}^\mathcal{T}$ \\ 
\hline
\pIn & $\psoc_{\pentry}, \Delta T^\mathcal{T}$ \\ 
\hline
\pOut & $\psoc_{\pexit}$ \\ 
\hline
\pA & $\Delta T^\mathcal{T} \ge 0 \And \psoc_{\pentry} \ge 0$\\ 
\hline
\pG &{$\begin{aligned}
\psoc_{\pentry -\pexit} &\in [\pcons_{\pmin\cdots\pmax}]\Delta T^\mathcal{T}\\
\psoc_{\pexit} &\in [0,100]
\end{aligned}$} \\
\hline
\end{tabular}
\end{center}

The constant hyperparameter, $\pcons_{\pminmax}^\mathcal{T}$, defines the range of power consumption depleting the battery during an instance of this task, which guarantees that the difference between the entry and exit state of charge will be proportional to the power consumption rate interval constant, $[\pcons_{\pmin\cdots\pmax}^\mathcal{T}]$, times the task duration input variable, $\Delta T^\mathcal{T}$.

\begin{table}[ht!]
\centering
\begin{tabular}{|cc|}
\hline
\pCST & $\pcons_{\pmin\cdots\pmax}^{\texttt{DSN,SBO,TCM\_h,TCM\_dv}},\pgen_{\pmin\cdots\pmax}^{\pCHRG}$ \\
\hline
\pIn & $\psoc_{\pentry}, \Delta T_{\texttt{DSN,SBO,TCM\_h,TCM\_dv}}$ \\
\hline
\pOut & $\psoc_{\pexit}^{\texttt{DSN,SBO,TCM\_h,TCM\_dv}}$ \\
\hline
\pA &{$\begin{aligned}
\Delta T_{\texttt{DSN,SBO,TCM\_h,TCM\_dv}} &\ge 0\\
\psoc_{\pentry} - \pcons_{\pmax}^{\pDSN} \Delta T_{\pDSN} &\ge 0
\end{aligned}$} \\
\hline
\pG &{$\begin{aligned}
\Delta \psoc^{\texttt{\pDSN}} &\in [\pcons_{\pmin\cdots\pmax}^{\pDSN}]\Delta T_{\pDSN}\\
\psoc_{\pexit}^{\pCHRG} - \psoc_{\pexit}^{\pDSN} &\in [\pgen_{\pmin\cdots\pmax}^{\pCHRG}]\Delta T_{\pCHRG}\\
\psoc_{\pexit}^{\pCHRG} &\le 100\\
\psoc_{\pexit}^{\pCHRG} - \psoc_{\pexit}^{\pSBO} &\in [\pcons_{\pmin\cdots\pmax}^{\pSBO}]\Delta T_{\pSBO}\\
\psoc_{\pexit}^{\pSBO} - \psoc_{\pexit}^{\pTCMh} &\in [\pcons_{\pmin\cdots\pmax}^{\pTCMh}]\Delta T_{\pTCMh}\\
\psoc_{\pexit}^{\pTCMh} - \psoc_{\pexit}^{\pTCMdv} &\in [\pcons_{\pmin\cdots\pmax}^{\pTCMdv}]\Delta T_{\pTCMdv}\\
\end{aligned}$} \\
\hline
\end{tabular}
\caption{5-step sequence power viewpoint composition}
\label{tbl:5-step-power-composition}
\end{table}

Composing the above for the 5-step scenario yields the contract in Table~\ref{tbl:5-step-power-composition}.
Notice the counter-intuitive assumption about the first step, \pDSN, requiring the scenario's initial state of charge to be greater than the worst-case consumption during the \pDSN~step; Pacti derived this assumption from the first step's contract guarantees. Furthermore, although each step contract guarantees an upper bound on the state of charge, Pacti's algebraic operations effectively captured the fact that this upper bound constraint is necessary for the \pCHRG~task and is otherwise implied for the subsequent power-consuming steps due to the chaining of the state of charge effects.

\subsubsection{Pacti contracts in the science \& communication viewpoint}
Qualitatively, this viewpoint is unaffected by \pCHRG~and \pTCM~tasks: their contracts reduce to a no-change guarantee that the science state variables on exit are equal to those on entry. For \pDSN, the range of downlink rate during the task instance depletes the onboard science data storage but leaves the cumulative science data acquired unaffected. For \pSBO, the science data generation rate range during the task instance increases both onboard science data storage and cumulative science data acquired. 

\begin{center}
\begin{tabular}{|cc|}
\multicolumn{2}{c}{Task template for $\mathcal{T}$\texttt{=DSN}}\\
\hline
\pCST & $[\prate_{\pmin\cdots\pmax}]$ \\
\hline
\pIn & $d_{\pentry}, c_{\pentry}, \Delta T^\mathcal{T}$ \\
\hline
\pOut & $d_{\pexit}, c_{\pexit}$\\
\hline
\pA & $\Delta T^\mathcal{T} \ge 0 \And d_{\pentry} \in [0,100]$\\
\hline
\pG & {$\begin{aligned}
d_{\pexit-\pentry} &\in [\prate_{\pmin\cdots\pmax}]\Delta T^\mathcal{T}\\
c_{\pexit-\pentry} &= 0\\
\end{aligned}$} \\
\hline
\end{tabular}
\end{center}

The constant hyperparameter, $\prate_{\pminmax}^\mathcal{T}$, defines the range of downlink rate draining the onboard science data storage during an instance of this task, which guarantees that the difference between the exit and entry data storage will be proportional to the downlink rate interval constant, $[\prate_{\pmin\cdots\pmax}^\mathcal{T}]$, times the task duration input variable, $\Delta T^\mathcal{T}$.

\begin{center}
\begin{tabular}{|cc|}
\multicolumn{2}{c}{Task template for $\mathcal{T}$\texttt{=SBO}}\\
\hline
\pCST & $\psgen_{\pmin\cdots\pmax}$ \\
\hline
\pIn & $d_{\pentry}, c_{\pentry}, \Delta T^\mathcal{T}$ \\
\hline
\pOut & $d_{\pexit}, c_{\pexit}$\\
\hline
\pA & {$\begin{aligned}
\Delta T^\mathcal{T} &\ge 0 \And c_{\pentry} \ge 0\\
d_{entry} &\in [0,100 - \psgen_{max} \Delta T_{SBO}]\\
\end{aligned}$} \\
\hline
\pG & {$\begin{aligned}
d_{\pexit} &\le 100\\
d_{\pexit - \pentry} &\in [\psgen_{\pmin\cdots\pmax}]\Delta T^\mathcal{T}\\
c_{\pexit - \pentry} &\in [\psgen_{\pmin\cdots\pmax}]\Delta T^\mathcal{T}\\
\end{aligned}$} \\
\hline
\end{tabular}
\end{center}

The constant hyperparameter, $\psgen_{\pminmax}^\mathcal{T}$, defines the range of science data generation rate accumulating in the onboard science data storage during an instance of this task, which guarantees that the difference between the exit and entry data storage (onboard and cumulative) will be proportional to the generation rate interval constant, $[\psgen_{\pmin\cdots\pmax}^\mathcal{T}]$, times the task duration input variable, $\Delta T^\mathcal{T}$.

Note that the \texttt{TCM\_h,TCM\_dv} tasks have trivial no-change contracts with the exit variables, $u,r$, equal to the corresponding entry variables. The overall science and communication viewpoint contract for the 5-step scenario yields the science and communication contract in Table~\ref{tbl:5-step-science-composition}.

\begin{table}[ht!]
\centering
\begin{tabular}{|cc|}
\hline
\pCST & $\prate_{\pminmax}^{\pDSN}, \psgen_{\pminmax}^{\pSBO}$ \\
\hline
\pIn & $d_{\pentry}, c_{\pentry}, \Delta T_{\texttt{DSN,SBO}}$ \\
\hline
\pOut & {$\begin{aligned}
d_{\pexit}^{\texttt{DSN,CHRG,SBO,TCM\_h,TCM\_dv}},\\
c_{\pexit}^{\texttt{DSN,CHRG,SBO,TCM\_h,TCM\_dv}}\\
\end{aligned}$} \\
\hline
\pA & {$\begin{aligned}
\Delta T_{\texttt{DSN,SBO}} &\ge 0\\
d_{\pentry} &\in [0,100]\\
\end{aligned}$} \\
\hline
\pG & {$\begin{aligned}
d_{\pexit} &\le 100\\
d_{\pentry - \pexit}^{\pDSN} &\in [\prate_{\pmin\cdots\pmax}]\Delta T_{\pDSN}\\
d_{\pexit - \pentry}^{\pSBO} &\in [\psgen_{\pmin\cdots\pmax}]\Delta T_{\pSBO}\\
c_{\pexit - \pentry}^{\pSBO} &\in [\psgen_{\pmin\cdots\pmax}]\Delta T_{\pSBO}\\
d_{\pexit}^{\texttt{TCM\_h,TCM\_dv}} &= d_{\pexit}^{\pSBO}\\
c_{\pexit}^{\texttt{TCM\_h,TCM\_dv}} &= c_{\pexit}^{\pSBO}\\
\end{aligned}$} \\
\hline
\end{tabular}
\caption{5-step sequence science \& communication viewpoint composition}
\label{tbl:5-step-science-composition}
\end{table}

\subsubsection{Pacti contracts in the navigation viewpoint}
Qualitatively, \pDSN~and \pCHRG~have similar impacts where the trajectory estimation uncertainty increases according to a noise range due to a change of spacecraft orientation performed during an instance of such tasks. Thanks to optimal measurements of the asteroid performed during this task, the onboard auto-navigation software can reduce the trajectory estimation uncertainty within some range of improvement. \pTCMh~has no impact on uncertainty. All three tasks leave the relative trajectory distance unchanged. Due to performing a long-duration change of velocity, the \pTCMdv~task injects additional trajectory estimation uncertainty proportional to a noise range; however, it reduces the relative trajectory distance proportional to an improvement range.

\begin{center}
\begin{tabular}{|cc|}
\multicolumn{2}{c}{Task template for $\mathcal{T} \in \{\texttt{DSN, CHRG}\}$}\\
\hline
\pCST & $\pnoise_{\pmin\cdots\pmax}$ \\
\hline
\pIn & $u_{\pentry}, r_{\pentry}$ \\
\hline
\pOut & $u_{\pexit}, r_{\pexit}$\\
\hline
\pA &$u_{\pentry} \in [0,100] \And r_{\pentry} \in [0,100]$\\
\hline
\pG & {$\begin{aligned}
r_{\pexit} = r_{\pentry} \And u_{exit} \le 100\\
u_{\pexit - \pentry} \in [\pnoise_{\pmin\cdots\pmax}]\\
\end{aligned}$} \\
\hline
\end{tabular}
\end{center}

The constant hyperparameter, $\pnoise_{\pmin\cdots\pmax}$, defines the range of trajectory estimation uncertainty noise injected in an instance of this task due to a single change of spacecraft orientation.

\begin{center}
\begin{tabular}{|cc|}
\multicolumn{2}{c}{Task template for $\mathcal{T}$\texttt{=SBO}}\\
\hline
\pCST & $\pimp_{\pmin\cdots\pmax}$ \\
\hline
\pIn & $u_{\pentry}, r_{\pentry}, \Delta T^\mathcal{T}$ \\
\hline
\pOut & $u_{\pexit}, r_{\pexit}$\\
\hline
\pA &$\Delta T^\mathcal{T} \ge 0 \And u_{\pentry} \le 100$\\
\hline
\pG & {$\begin{aligned}
r_{\pexit} = r_{\pentry} \And u_{exit} \in [0,100]\\
u_{\pexit - \pentry} \in [\pimp_{\pmin\cdots\pmax}] \Delta T^\mathcal{T}\\
\end{aligned}$} \\
\hline
\end{tabular}
\end{center}

The constant hyperparameter, $\pimp_{\pmin\cdots\pmax}$, defines the range of trajectory estimation uncertainty improvement during an instance of this task due to onboard autonomous navigation calculations, which guarantees that the difference between the exit and entry uncertainty will be proportional to the improvement rate interval constant\footnote{An improvement interval with a negative lower bound corresponds to the possibility of a navigation trajectory deterioration.}, $[\pimp_{\pmin\cdots\pmax}]$, times the task duration input variable, $\Delta T^\mathcal{T}$.

Note that the \pTCMh~task has a trivial no-change contract with the exit variables, $u,r$, equal to the corresponding entry variables. 

\begin{center}
\begin{tabular}{|cc|}
\multicolumn{2}{c}{Task template for $\mathcal{T}$\texttt{=TCM\_dv}}\\
\hline
\pCST & $\pimp_{\pmin\cdots\pmax}, noise_{\pmin\cdots\pmax}$ \\
\hline
\pIn & $u_{\pentry}, r_{\pentry}, \Delta T^\mathcal{T}$ \\
\hline
\pOut & $u_{\pexit}, r_{\pexit}$\\
\hline
\pA &$\Delta T^\mathcal{T} \ge 0 \And u_{\pentry} \in [0,100] \And r_{entry} \le 100$\\
\hline
\pG & {$\begin{aligned}
r_{\pexit} \ge 0 &\And u_{\pexit} \in [0,100]\\
r_{\pexit - \pentry} &\in [\pimp_{\pmin\cdots\pmax}] \Delta T^\mathcal{T}\\
u_{\pexit - \pentry} &\in [\pnoise_{\pmin\cdots\pmax}] \Delta T^\mathcal{T}\\
\end{aligned}$} \\
\hline
\end{tabular}
\end{center}

The constant hyperparameter, $\pimp_{\pmin\cdots\pmax}$, defines the range of relative trajectory progress improvement during an instance of this task due to onboard autonomous navigation calculations, which guarantees that the difference between the exit and entry progress will be proportional to the improvement rate interval constant, $[\pimp_{\pmin\cdots\pmax}]$, times the task duration input variable, $\Delta T^\mathcal{T}$. The constant hyperparameter, $\pnoise_{\pmin\cdots\pmax}$, defines the range of trajectory estimation uncertainty degradation during an instance of this task due to velocity change being performed, which guarantees that the difference between the exit and entry uncertainty will be proportional to the noise interval constant, $[\pnoise_{\pmin\cdots\pmax}]$, times the task duration input variable, $\Delta T^\mathcal{T}$. 

Composing the above for the 5-step scenario yields the navigation contract in Table \ref{tbl:5-step-nav-composition}.

\begin{table}[ht!]
\centering
\begin{tabular}{|cc|}
\hline
\pCST & $\pnoise_{\pminmax}^{\texttt{DSN,CHRG,TCM\_dv}}, \pimp_{\pminmax}^{\texttt{SBO,TCM\_dv}}$ \\
\hline
\pIn & $u_{\pentry}, r_{\pentry}, \Delta T_{\texttt{SBO,TCM\_dv}}$ \\
\hline
\pOut & {$\begin{aligned}
u_{\pexit}^{\texttt{DSN,CHRG,SBO,TCM\_h,TCM\_dv}},\\
r_{\pexit}^{\texttt{DSN,CHRG,SBO,TCM\_h,TCM\_dv}}\\
\end{aligned}$} \\
\hline
\pA & {$\begin{aligned}
\Delta T_{\texttt{SBO,TCM\_dv}} \ge 0\\
u_{\pentry} \in [0,100] \And r_{\pentry} &\in [0,100]\\
\end{aligned}$} \\
\hline
\pG & {$\begin{aligned}
u_{\pexit - \pentry}^{\pDSN} &\in [\pnoise_{\pmin\cdots\pmax}^{\pDSN}]\\
r_{\pexit}^{\pDSN} &= r_{\pentry}\\
u_{\pexit}^{\pCHRG} - u_{\pexit}^{\pDSN} &\in [\pnoise_{\pmin\cdots\pmax}^{\pCHRG}]\\
u_{\pexit}^{\pCHRG} &\le 100\\
u_{\pexit}^{\pCHRG} - u_{\pexit}^{\pSBO} &\in [\pimp_{\pmin\cdots\pmax}^{\pSBO}] \Delta T_{\pSBO}\\
u_{\pexit}^{\pSBO} \ge 0 &\And
r_{\pexit}^{\pSBO} = r_{\pexit}^{\pCHRG}\\
u_{\pexit}^{\pTCMh} = r_{\pexit}^{\pSBO} &\And r_{\pexit}^{\pTCMh} = r_{\pexit}^{\pSBO}\\
u_{\pexit}^{\pTCMdv} - u_{\pexit}^{\pTCMh} &\in [\pnoise_{\pmin\cdots\pmax}^{\pTCMdv}] \Delta T_{\pTCMdv}\\
r_{\pexit}^{\pTCMdv} - r_{\pexit}^{\pTCMh} &\in [\pimp_{\pmin\cdots\pmax}^{\pTCMdv}] \Delta T_{\pTCMdv}\\
\end{aligned}$} \\
\hline
\end{tabular}
\caption{5-step sequence navigation viewpoint composition}
\label{tbl:5-step-nav-composition}
\end{table}

%% file: schedule-analysis.tex
\subsection{Schedulability analysis}
\label{mission:schedule-analysis}
Our schedulability analysis methodology reflects separating design concerns from operation concerns. Design concerns correspond to the capability characteristics of each task:
\begin{itemize}
    \item range of power consumption for each of the tasks \texttt{DSN,SBO,TCM\_h,TCM\_dv};
    \item range of power generation for the \texttt{DSN} task;
    \item min,max range of downlink speed for the \texttt{DSN} task;
    \item range of science data acquisition rate for the \texttt{SBO} task;
    \item range of trajectory estimation uncertainty noise injection for each of \texttt{DSN,CHRG,TCM\_dv} tasks;
    \item range of trajectory estimation uncertainty improvement due to optimal small body measurements for the \texttt{SBO} task; and
    \item range of relative trajectory progress for the \texttt{TCM\_dv} task.
\end{itemize}

We define these capability characteristics as contract hyperparameters, as discussed in Section \ref{sec:scenario-tasks}. On the other hand, we define operational requirements as constraints on entry/exit variables:

\begin{itemize}
    \item Minimum battery state of charge: 60-90\%
    \item Minimum task duration for each step: 10-50 seconds
    \item Initial science data volume: 60-100\%
    \item Initial trajectory estimation uncertainty: 40-90\%
\end{itemize}

\begin{table*}
\begin{tabular}{|c|p{6cm}|p{6cm}|}
\hline
Scenario
& Pacti operations for scenario generation 
& Pacti operations for schedulability analysis
\\
\hline
5-step & 
compose: 2016, up to 22 constraints, 12 variables\newline 
merge: 1680, up to 44 constraints, 23 variables\newline
total time: 16.3 seconds
& merge: 181,988, up to 81 constraints, 35 variables\newline
admissible solutions: 401 out of 20,000 combinations\newline
total time: 3 minutes, 9.6 seconds\\
\hline
20-step & 
compose: 10,200, up to 187 constraints, 95 variables\newline
merge: 8,000, up to 44 constraints, 23 variables\newline
total time: 97.8 seconds
& merge: 781,331, up to 275 constraints, 125 variables \newline
admissible solutions: 244 out of 20,000 combinations\newline
total time: 4 minutes, 11 seconds\\
\hline
\end{tabular}
\caption{Scenario generation (100 hyperparameter samples) \& schedulability analysis (100 operational requirement samples)}
\label{tab:results}
\end{table*}

\begin{figure*}
\centering
\begin{subfigure}{0.47\textwidth}
\includegraphics[width=0.9\linewidth]{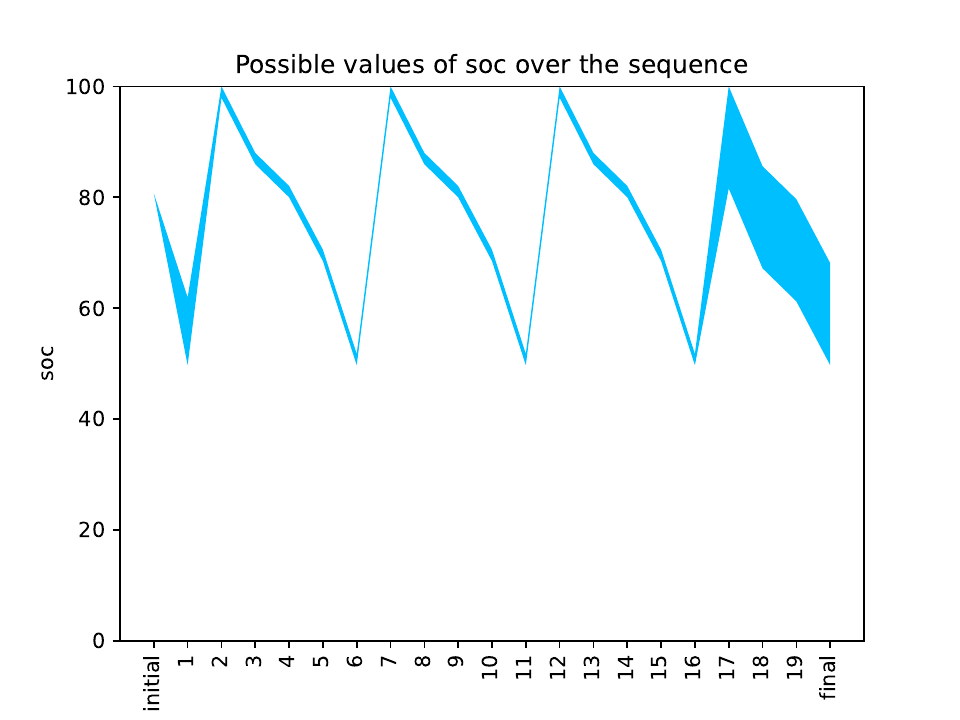}
\caption{Low-variability schedule with high min. soc levels}
\label{fig:results-a}
\end{subfigure}
\begin{subfigure}{0.47\textwidth}
\includegraphics[width=0.9\linewidth]{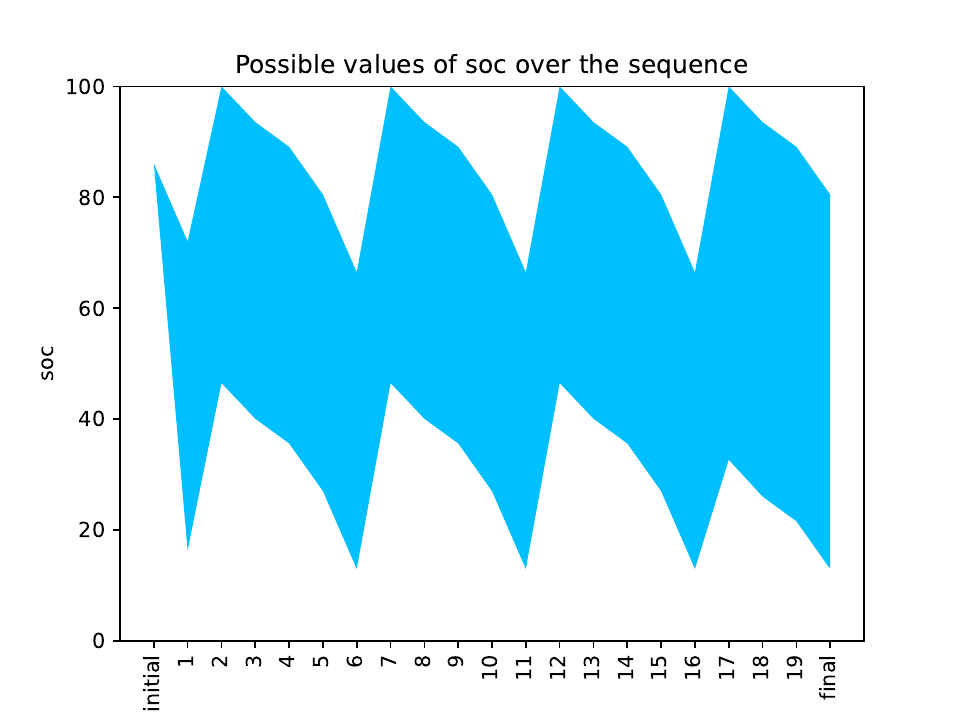}
\caption{High-variability schedule with potentially low soc levels}
\label{fig:results-b}
\end{subfigure}
\caption{Examples of schedulability analysis results}
\end{figure*}

Methodologically, we defined schedulability as the compatibility between a schedule (based on a given choice of capability hyperparameters) and a set of operational requirements (based on a given choice of values for entry/exit variables). We applied the Latin hypercube statistical sampler to generate multiple combinations of scenarios and operational requirements as summarized in Table \ref{tab:results} using a Windows 10 workstation powered by an AMD Threadripper Pro 3955WX processor with 16 cores and 128GB RAM running Ubuntu 20.04 under Windows 10's WSL2\footnote{For details about performance measurement and API statistics, see \url{https://github.com/pacti-org/pacti-instrumentation}.}. For scenario generation, we sampled 200 distributions to generate mean and deviation for specifying the range of each of the 12 capability hyperparameters. The second column shows the statistics for producing a short 5-step scenario and a long 20-step scenario given such a sample. For varying operational requirements, we generated 100 random values within predefined ranges of requirement constraints. We computed schedulability using Pacti's merge operation for all combinations of 200 scenarios and 100 operational requirements. The third column shows the statistics for this schedulability analysis. The scarcity of admissible solutions (i.e., less than 1\%) and the efficiency of schedulability analysis\footnote{Over 100 combinations per second (5-step scenario) and over 25 combinations per second (20-step scenario) using up to 32 concurrent jobs.} demonstrates the usefulness of Pacti for rapid exploration of design and operational constraints.

Pacti's API provides additional capabilities to get useful insights into admissible schedules. For example,
Figures \ref{fig:results-a} and \ref{fig:results-b} show two results of the visualizing the bounds of battery state-of-charge at the entry and exit of each step in the schedule using Pacti's \texttt{get\_variable\_bounds()} API. Note that this range visualization is qualitatively different from a simulation timeline: a single value over time. These figures also illustrate a subtle aspect of Pacti's polyhedral contract algebra where the effect of composing the sequence step contracts results in relaxing the guarantees \cite[Sec. 4]{pacti}, broadening the possible exit variable ranges since the final variables are unconstrained. Conversely, the middle section of the scenario shows greater precision in the bound calculations since the subsequent contracts force constraints on the exit variables, thereby preventing their relaxation. Aside from the subtleties of contract relaxation, these two figures show a stark contrast between different scenario characteristics and operational requirements combinations. Such differences would compellingly motivate cross-validating the Pacti contract approximations with appropriate simulation models to get additional insights into these differences.

\begin{figure}
\centering
\includegraphics[width=\columnwidth]{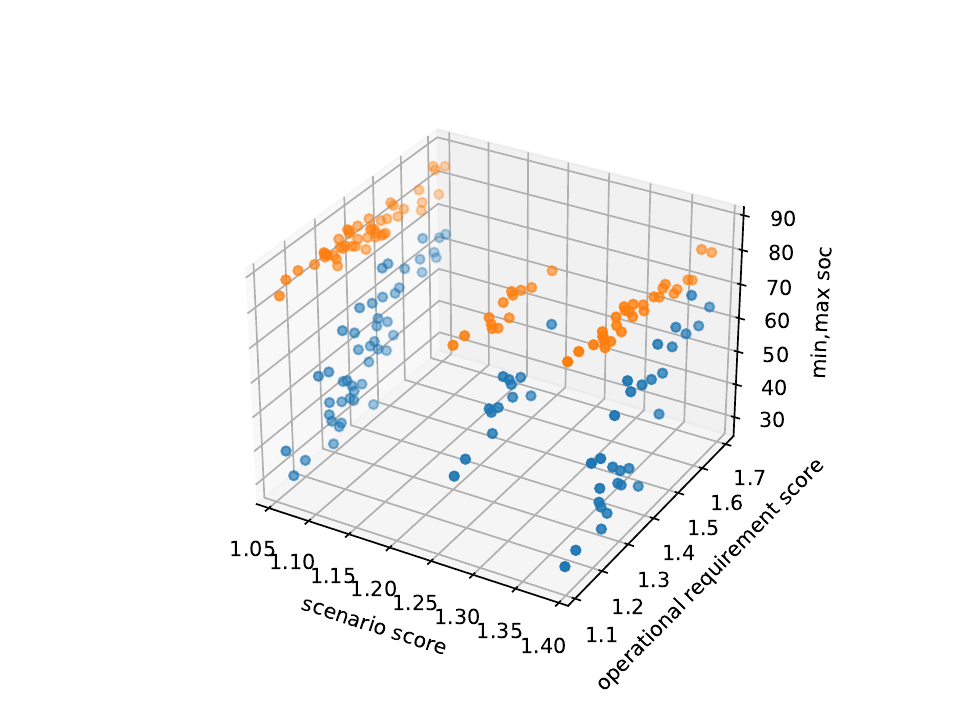}
\caption{Score-based visualization of the min (blue) and max (orange) battery state of charge over all admissible schedules}
\label{fig:min-max-scoring}
\end{figure}

With Pacti's \texttt{optimize()} API, we computed the minimum and maximum values of a linear optimization metric, the average of all states of charges at the end of each step, and plotted these results in Figure \ref{fig:min-max-scoring} by scoring each admissible schedule w.r.t the scenario and operational requirement. For scenario scoring, we took the average of adding/subtracting all viewpoint-specific positive/negative capabilities: generation vs. consumption for the power viewpoint, downlink speed vs. observation rate for the science viewpoint, and noise vs. improvement for the navigation viewpoint.
For operational requirement scoring, we averaged all constraints since the difficulty of achieving them increases with their magnitude. The clustering of admissible schedules suggests that designers could get more insight by performing this scoring on specific viewpoints instead of combining them as was done here.

%% file: aircraft.tex
\begin{figure}
    \begin{center}
        \includegraphics[width=0.8\textwidth]{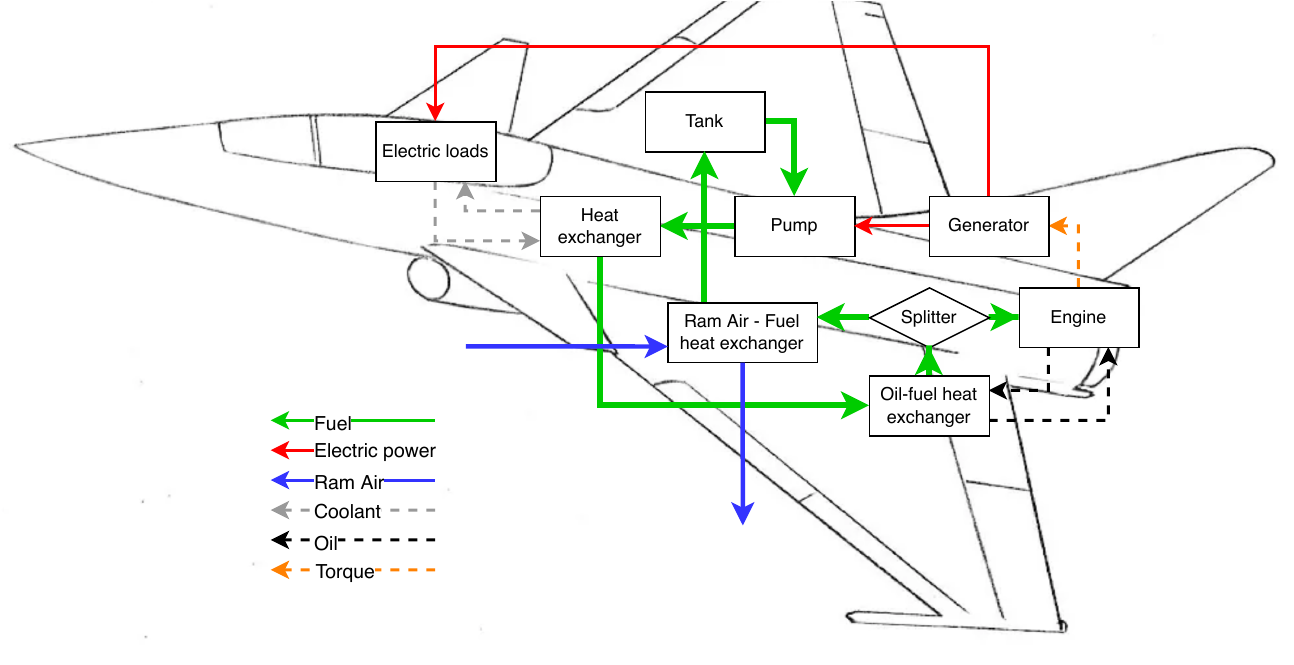}
        \caption{A pictorial representation of the interactions between the fuel system, the thermal management system and the electrical system.}
        \label{fig:aircraft_high_level_picture}
    \end{center}
\end{figure}

After having considered an application of contracts and Pacti to the design of space missions,
in this section we will consider their application in the design the design of the thermal management system for aircraft.

\subsection{Description of the thermal management system}

The dependencies of a few sub-systems of a prototypical aircraft are shown in Figure~\ref{fig:aircraft_high_level_picture}. The propulsion system is represented by an engine which transforms the chemical energy stored in the fuel into thrust and mechanical power. The mechanical power is transformed into electric power by a generator. The generation of mechanical power and electric power are both inefficient processes that generate heat. Fuel must be delivered from the tank to the engine at a certain rate which depends on the required thrust level. A fuel pump serves this purpose by moving fuel along a circuit using, in our example, electric power from the generator. An aircraft has several other electric loads, such as actuators and flight computers that, not being 100\% efficient, generate heat as well. The heat generated by the electric loads, the generator, and the engine can be used to maintain a desirable fuel temperature which would otherwise be too low at high-altitudes to operate the engine efficiently. Thus, heat exchangers are used to transfer heat to the fuel before reaching the engine. Some fuel is returned to the tank for two reasons. First, the fuel flow rate must be regulated to absorb heat from the components on the aircraft while also maintaining the fuel temperature at the engine inlet within prescribed bounds. Secondly, the fuel in the tank must also be maintained at a desirable temperature (definitely above the freezing point and below the burning point) which can be achieved by returning hot fuel to the tank. The returned fuel could be too hot though, and its temperature may have to be reduced by rejecting some heat through a heat exchanger with the outside air. 

We are interested in the problem of designing the system so that the temperature at the engine inlet and in the tank are always kept within acceptable ranges over the entire flight envelope. The design has to be robust with respect to uncertainties in the generated heat, and component tolerances. Ideally, the result of the design phase is a set of specifications for the parameters of the components in the system. The wider the range allowed for these parameters, the wider the set of components to choose from. Also, in general, larger tolerances correspond to less expensive manufacturing processes. The key variables we consider are the following: the flight regime is defined by the pair $(alt, thrust)$ of the flight altitude and thrust level, respectively, and the operating point is defined by the pair $(\dot{m}_{in},\dot{m}_a)$ of the fuel flow rate imposed by the pump and the air flow rate used to cool down the fuel returning to the tank, respectively. The system-level specification $Spec$ defines the range of allowed values for the temperature at the engine $T_e$, and at the tank inlet $T_{out}$.  Thus, we can define two problems that we wish to address: 

\begin{problem}[Analysis]
\label{problem:analysis}
Given a range for $(alt,thrust)$, and given component models and associated uncertainties, check whether a pair $(\dot{m}_{in},\dot{m}_a)$ satisfies system level specification $Spec$.
\end{problem}

\begin{problem}[Optimization]
\label{problem:optimization}
Given a feasible pair $(\dot{m}_{in},\dot{m}_a)$, and given a map that associates a cost to a component as a function of tolerances and inefficiencies, find the optimal distribution of these parameters among the components of the system such that total cost is minimized.
\end{problem}

\begin{figure}
    \begin{center}
        \includegraphics[width=0.8\textwidth]{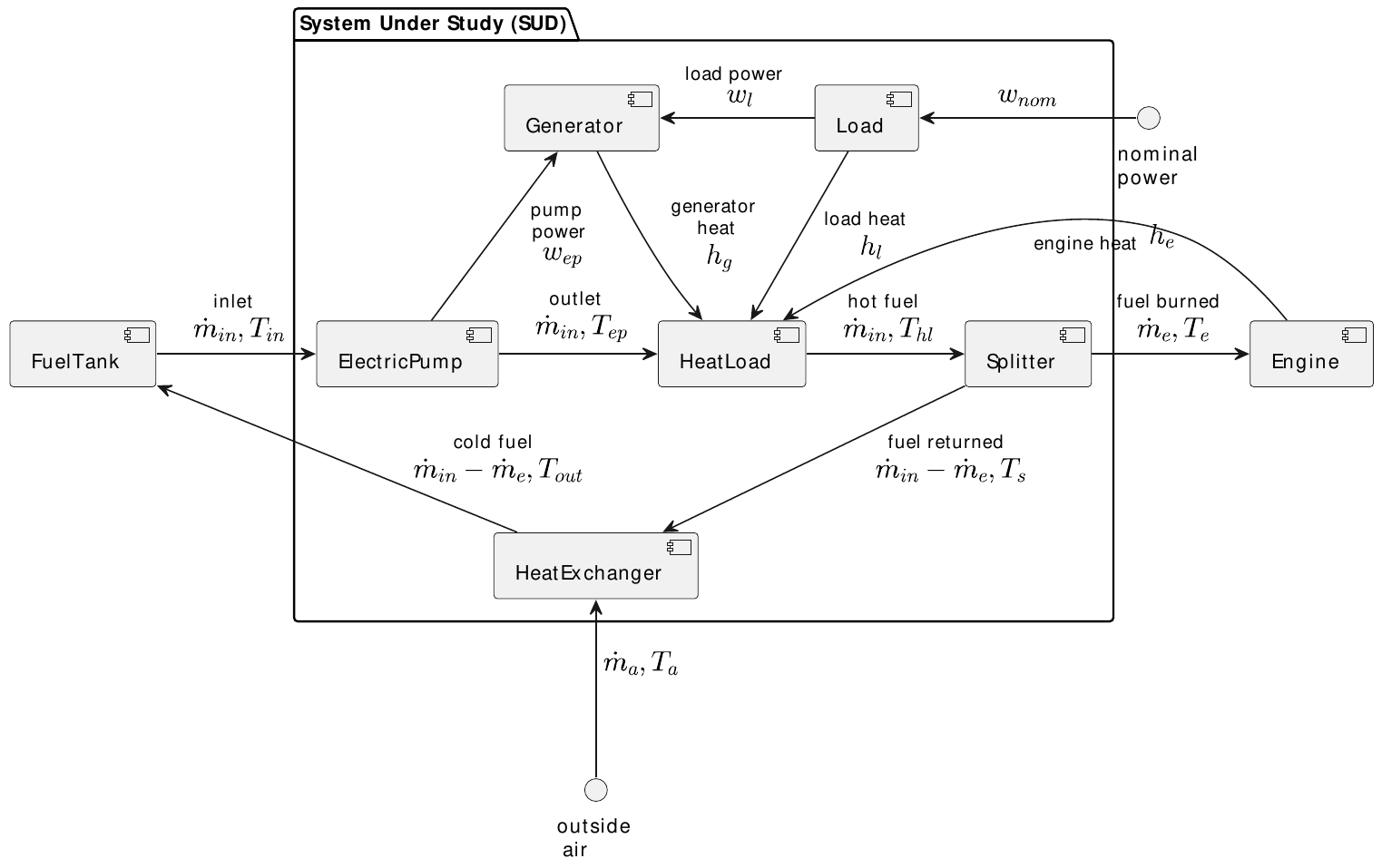}
        \caption{Block diagram of the System Under Study}
        \label{fig:aircraft_ftms}
    \end{center}
\end{figure}

In order to address these two problems, we abstract the system into the block diagram shown in Figure~\ref{fig:aircraft_ftms}. This block diagram focuses on the fuel system only, and abstracts the other subsystems at their interfaces with the fuel system. This block diagram shows how the system works: fuel flows from the tanks to the engines using a pump; the heat generated by electrical and electronic devices, and the heat generated by the engines, is transferred to the fuel to increase its temperature; some hot fuel is burned by the engines, while some is returned to the tank; the fuel that returns back may become too hot, and its temperature may have to be decreased through a heat exchanger that uses external air as cold fluid. 

The electric pump determines the fuel flow rate $\dot{m}_{in}$ at the input of the system which is equal to the flow rate at the output of the electric pump. The temperature of the fuel at the inlet of the pump is $T_{in}$, while the temperature at its output is $T_{ep}$. A heat exchanger collects heat $h_g$ from the electric power generator, $h_l$ from the electric load (representing a lumped model of the electric distribution system and various electric loads), and $h_e$ from the engines. These three heat sources increase the temperature of the fuel to $T_{hl}$. The fuel is then split into two paths: the engine burns fuel at a rate $\dot{m}_e$, while the remaining flow returns to the tank at a rate $\dot{m}_{in} - \dot{m}_e$. Finally, the fuel is cooled down through a fuel-air heat exchanger which brings the temperature $T_{out}$ to an acceptable value. 

\subsection{Leveraging contracts for analysis and optimization}
One way to tackle Problems~\ref{problem:analysis} and~\ref{problem:optimization} is to develop a simulation model, and sample its parameters in a certain range. For example, assume the electric load in Figure~\ref{fig:aircraft_ftms} requires a nominal power $w_l$, but the actual power requirement is in the range $[(1-\epsilon_l)\cdot w_l, (1+\epsilon_l)\cdot w_l]$, where the tolerance factor $\epsilon_l$ models several sources of uncertainty. Then the model would be simulated for different values of $w_l$ in this range. The data gathered by these simulation runs is used to create the response surfaces for $(\dot{m}_{in},\dot{m}_a)$ and for total cost. Samples that violate the assumptions of any of the components, or the system specification are simply rejected as invalid. As the number of parameters that can span many values grows, the number of simulation runs grows geometrically. 

We leverage a contract-based framework to represent explicitly the assumptions and the guarantees for each component, and compute algebraically  the temperature ranges for $T_e$ and $T_{out}$. We represent the system specification as a contract $Spec$ where the inputs include the flight regime, the temperature of the fuel in the tank, and the nominal power requirement, and the outputs are the temperature of the fuel $T_e$ at the engine inlet and the temperature going back to the tank $T_{out}$. Each component in Figure~\ref{fig:aircraft_ftms} is modeled by its own contract where the guarantee captures a range of implementations by defining bounds on the values of the component's outputs. We leverage the ability of Pacti to compute an explicit representation of the composition of contracts to derive a single contract for the system under study (SUD). This contract provides us with the entire range of possible values for the two key variables $T_e$ and $T_{out}$. We also use Pacti's ability to check whether the SUD refines the specification $Spec$. These features are also leveraged during optimization where Pacti is called in the optimization loop to compute the bounds for $T_e$ and $T_{out}$ which are  used in the computation of the cost function and constraint violations. 

\subsection{Modeling the system using polyhedral assume-guarantee contracts} 
Consider a connection between two components in the fuel circuit. Let $\dot{m}$ (in \unit{\kilogram\per\second}) denote the fuel flow rate, and $T$ (in \unit{\kelvin}) denote the fuel temperature. The heat rate through such connection is $\dot{m}\cdot C_f \cdot T$, where $C_f$ is the specific heat of the jet fuel ($0.2$ \unit{\kJ\per\kilogram\per\kelvin}). The heat rate is an important quantity in this model because balancing heat while satisfying temperature and fuel flow rate constraints is the key problem in this application. However, this term involves the product of two key quantities (the fuel flow rate and the temperature) that are also involved in other constraints. If these quantities are both considered variable in our analysis, then the heat rate becomes a non-linear term which, and the model would fall outside of the polyhedral constraint required to perform analysis using Pacti. One of the two sets of variables (either fuel flow rates or temperatures) must be treated as as a set of parameters which are fixed to a constant value. A natural choice in this case study is to consider the fuel temperatures at different points in the system as variables because they must obey operational constraints. Specifically, the temperate at the engine and tank inlet must lie within acceptable ranges. 

The top-level specification of the system under study can be defined as a contract $Spec$ with the set of input variables $I_{Spec} = \{ T_{in}$, $T_a, w_{nom} \}$, representing the input temperature from the tank, the air temperature, and the nominal power requirement from the electrical components on the aircraft. The set of output variables is $O_{Spec}=\{\ T_e,T_{out}\}$. The specification contract is $Spec =(I_{Spec},O_{Spec},A_{Spec},G_{Spec})$, where the assumption $A_{Spec}$ specifies bounds around nominal values of the input temperature, the air temperatures, and  the power requirement, while the guarantee $G_{Spec}$ specifies acceptable ranges of the temperature at the engine and at the output.

\begin{figure}
    \begin{center}
        \includegraphics[width=\textwidth]{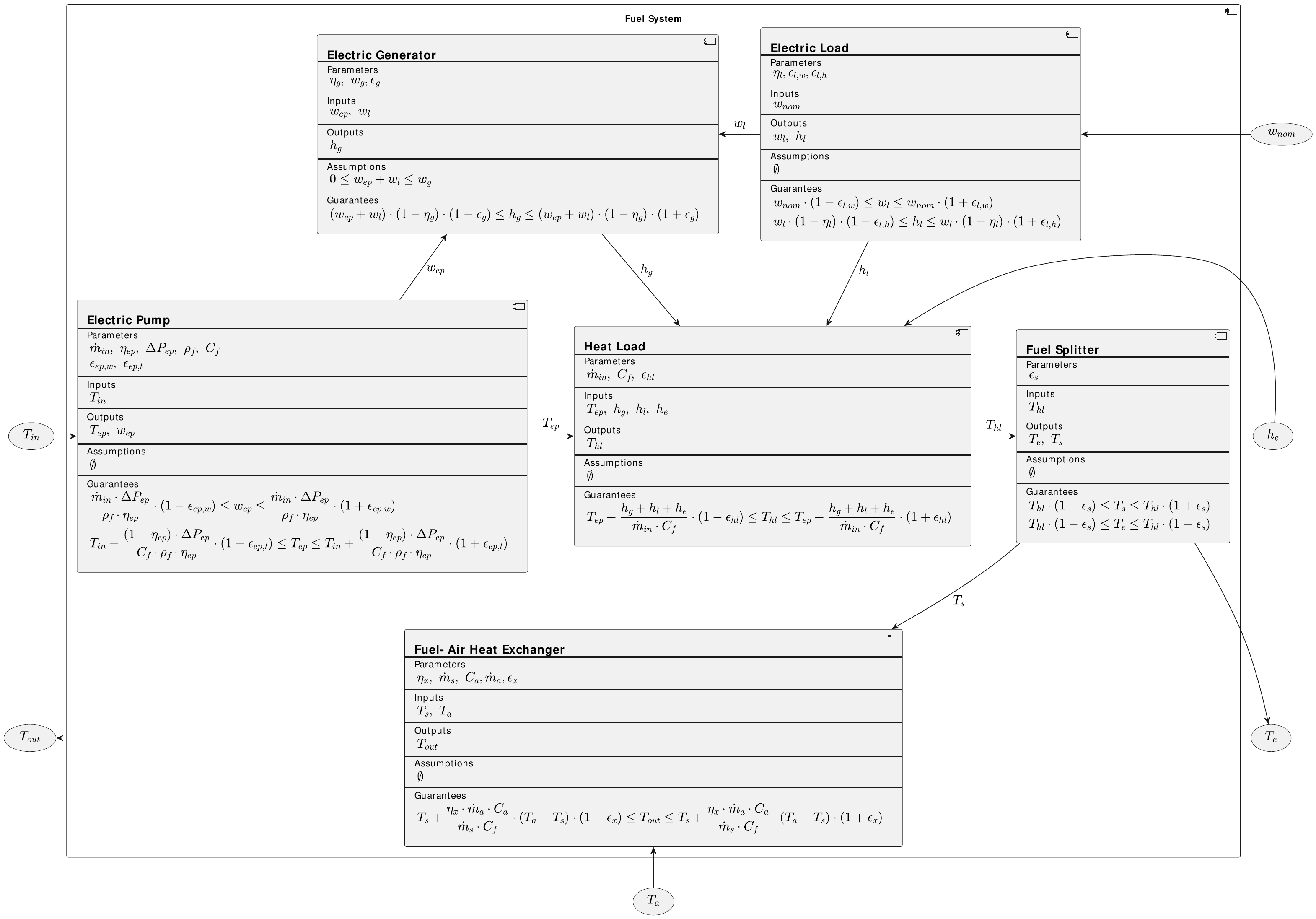}
        \caption{Contract-based model of the SUD.}
        \label{fig:pacti_aircraft_const_mdot}
    \end{center}
\end{figure}

The complete model of the contracts used to define the System Under Study (SUD) is shown in Figure~\ref{fig:pacti_aircraft_const_mdot}.
The electric pump has an inlet, an outlet, and an electrical interface. The pressure difference between the inlet and the outlet is $\Delta P_{ep}$ which we assume to be $6.9$ \unit{MPa}. The electric power required by the pump is $w_{ep} = \frac{\dot{m}_{in} \cdot \Delta P_{ep}}{\rho_f \cdot \eta_{ep}}$ where $\rho_f=800$ \unit{\kilogram\per\meter^3} is the density of the fuel, and $\eta_{ep}$ is the efficiency of the pump which we assume to be $0.6$. Some power, specifically $w_{ep} \cdot (1 - \eta_{ep})$, is transformed into heat which is absorbed by the fluid going through the pump. The pump increases the temperature of the fuel by $\frac{(1 - \eta_{ep})\cdot \Delta P_{ep}}{C_f \cdot \rho_f \cdot \eta_{ep}}$ \unit{\kelvin}. We use two parameters $\epsilon_{ep,w}$ and $\epsilon_{ep,t}$ to capture the power and temperature uncertainties which can also be seen as characterizing a space of possible implementations for the pump. This is also a general modeling pattern that we use throughout the development of the component models for this use case.

The heat load on the main fuel line to the engine increases the fuel temperature by $\frac{h_g + h_l + h_e}{\dot{m}_{in} \cdot C_f}$ \unit{\kelvin}, while the fuel-air heat exchanger decreases the temperature of the fuel returning to the tank by $\frac{\eta_x \cdot \dot{m}_a \cdot C_a}{\dot{m}_s \cdot C_f} \cdot (T_s - T_a)$\footnote{The heat exchanger equation can be found in \url{https://web.mit.edu/16.unified/www/FALL/thermodynamics/notes/node131.html}.} \unit{\kelvin}, where $\eta_x$ is the efficiency of the heat exchanger which we assume to be $0.6$ (and which models several factors including the size of the exchanger), and $C_a$ is the specific heat of the outside air which we assume to be $1$ \unit{\kJ\per\kilogram\per\kelvin}). 

\begin{figure}
    \begin{center}
        \includegraphics[width=\textwidth]{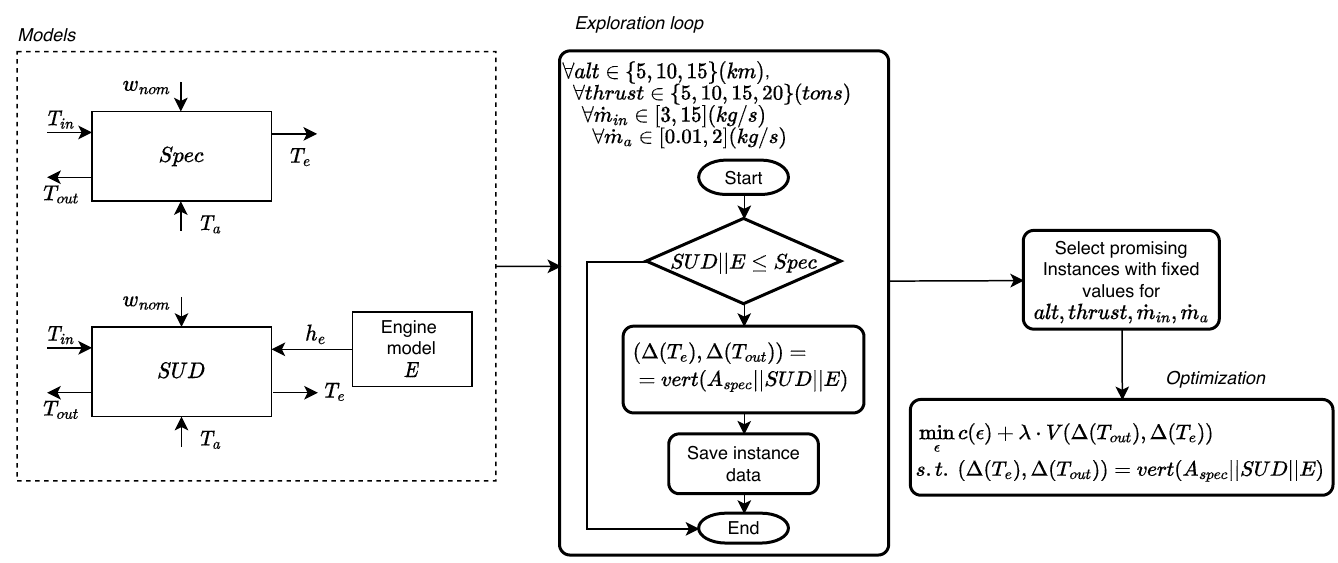}
        \caption{Exploration and optimization methodology.}
        \label{fig:exploration-optimization}
    \end{center}
\end{figure}

\subsection{Analysis and design space exploration}
We use these models to explore and optimize the SUD according to the methodology shown in Figure~\ref{fig:exploration-optimization}. The exploration sweeps over three flight altitudes ($5$, $10$, and $15$ \unit{\km}), and four levels of engine thrust ($5,000$, $10,000$, $15,000$, and $20,000$ \unit{\kilogram}). We map the flight altitude to a nominal value for the air temperature $T^*_a$ according to the model described in \cite{atmospheric1976us}, and we map thrust to a nominal value for the fuel flow as $\dot{m}_e = 0.7 \cdot thrust / 3600$. We also fix the nominal power requirement to $w^*_{nom} = 140$ \unit{\kW}, and the nominal fuel temperature in the tank to $T^*_{in} = 288$ \unit{\kelvin}. For each operating point, the exploration loop sweeps over a range of fuel flow $\dot{m}_{in}$ and air-flow $\dot{m}_a$. The ranges were selected to cover many possible implementations of pumps and heat exchangers. 

After selecting these parameters, we use Pacti to compute the contract of the system $SUD$ as the composition of the the electric pump, electric generator, electric load, heat load, fuel splitter, and heat exchanger (as shown in Figure~\ref{fig:pacti_aircraft_const_mdot}).
We then compose the $SUD$ with an engine model $E$ which is used to define the engine heat $h_e$ as a function of the parameter $\dot{m}_e$. The nominal engine model is $h_e = k_e \cdot \dot{m}_e$ with $k_e = 5,000$ \unit{\joule\per\kilogram}. The resulting system $SUD \parallel E$ must refine the specification contract $Spec$. 

We define $A_{Spec} \equiv \overline{T_{in}} \leq T_{in} \leq \underline{T_{in}} \wedge \overline{T_{a}} \leq T_{a} \leq \underline{T_{a}} \wedge \overline{w_{nom}} \leq w_{nom} \leq \underline{w_{nom}}$. The bounds $\overline{T_{in}}$, $\underline{T_{in}}$, $\overline{T_{a}}$, and $\underline{T_{a}}$ are defined as 2\% tolerances around their nominal values, while $\overline{w_nom}$ and $\underline{w_{nom}}$ are defined as 5\% tolerances around their nominal values. The guarantee of this system is $G_{Spec} \equiv \underline{T_{e}} \leq T_{e} \leq \overline{T_{e}} \wedge T_{in} - \Delta_{t} \leq T_{out} \leq T_{in} + \Delta_{t}$. The output temperature needs to be close enough to the input temperature to maintain the temperature of the fuel in the tank approximately constant. We use $\Delta_t = 10$ \unit{\kelvin} in our analysis, and we fix $\underline{T_{e}} = 300$ \unit{\kelvin} and $\overline{T_{e}} = 330$ \unit{\kelvin}. 

If $SUD \parallel E \leq Spec$, then we compute the actual temperature bounds at the engine and at the output. We leverage the unique capability of Pacti to compute and explicit polyhedral representation of $SUD \parallel E \parallel A_{Spec}$ where, with abuse of notation, we have denote by $A_{Spec}$ the contract $(True,A_{Spec})$. Pacti can then compute the extreme vertices of $SUD \parallel E \parallel A_{Spec}$ which correspond to the actual ranges $\Delta(T_e)$ and $\Delta(T_{out})$ of the output variables $T_e$ and $T_{out}$, respectively.

\begin{figure}
    \begin{center}
        \includegraphics[width=\textwidth]{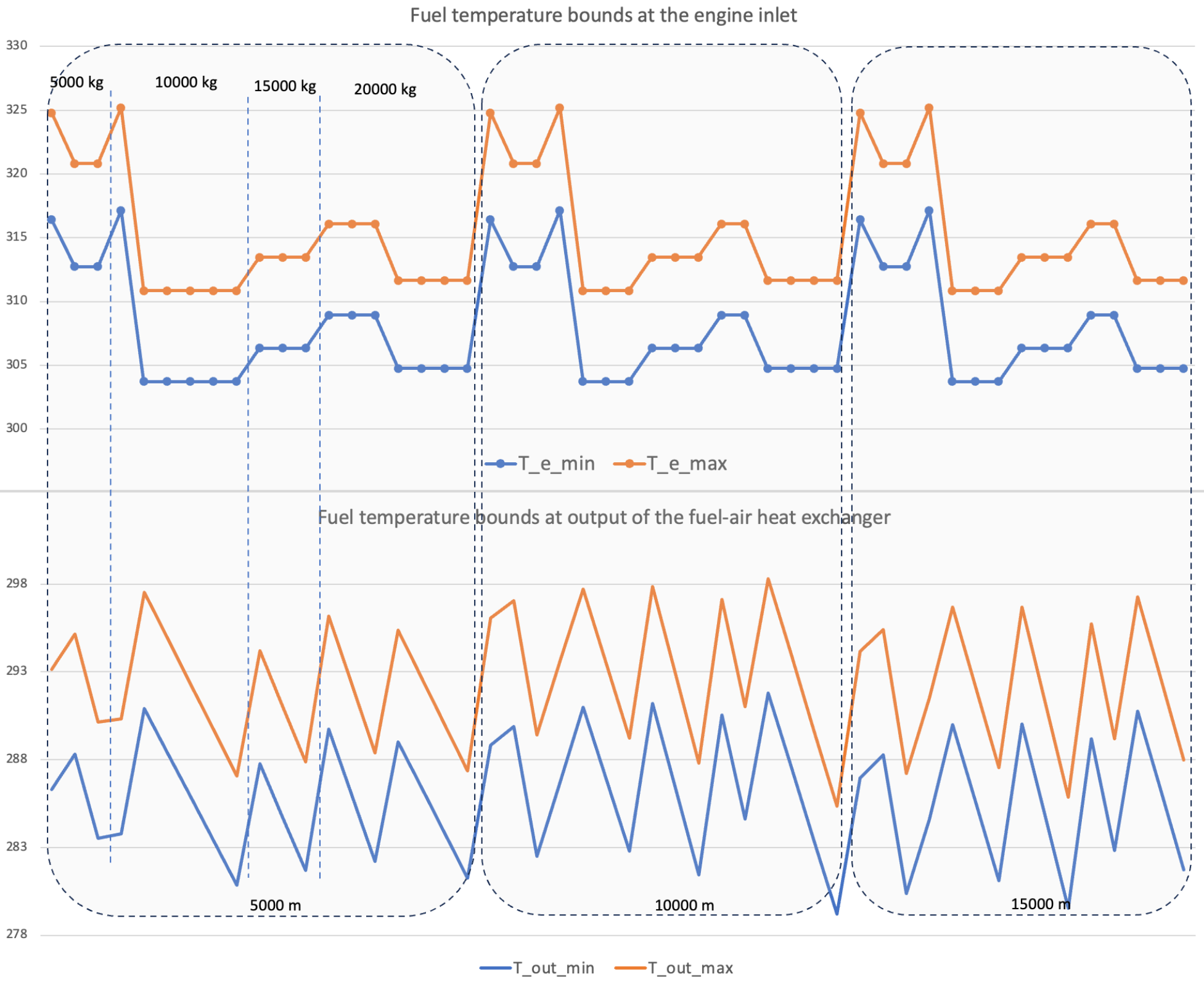}
        \caption{The $\Delta(T_e)$ and $\Delta(T_{out})$ obtained by the exploration loop. The $x$-axis represents successive instances. For each altitude and thrust level, $\dot{m}_{in}$ increases from left to right, and for each $\dot{m}_{in}$, $\dot{m}_a$ increases from left to right as well.}
        \label{fig:t_e_bounds}
    \end{center}
\end{figure}

Figure~\ref{fig:t_e_bounds} shows the temperature bounds computed by the exploration loop. As expected, the temperature bounds at the engine do not depend on $\dot{m}_a$. Higher values of $\dot{m}_{in}$ result in lower values of temperate bounds at the engine, while higher values of $\dot{m}_a$ result in lower values of temperature bounds at the output.

\begin{figure}
    \centering
    \begin{subfigure}{0.5\textwidth}
        \centering
        \includegraphics[width=0.9\textwidth]{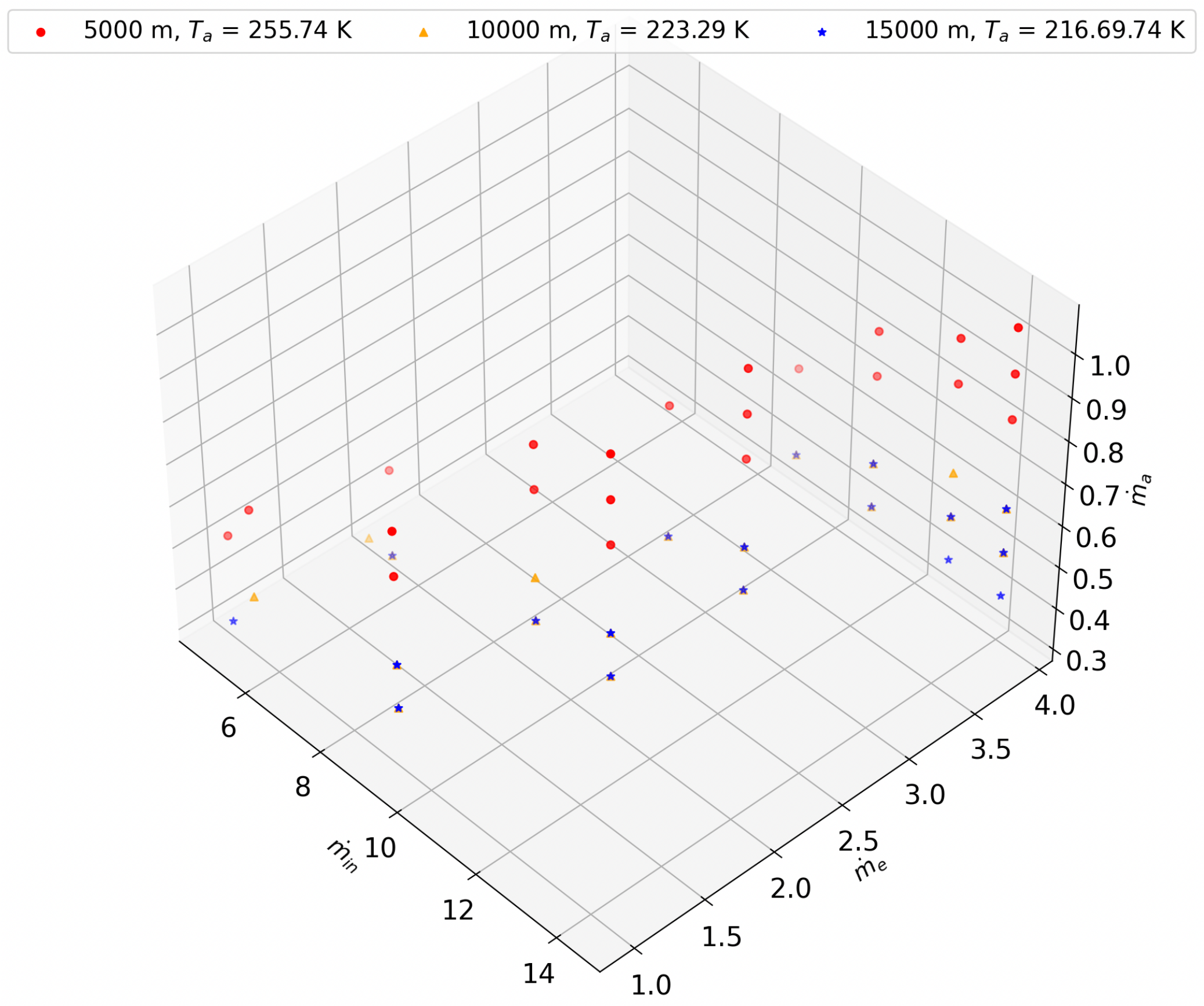}
        \caption{Valid fuel and air flow values for a system operating at different altitudes with a fixed heat exchanger.}
    \end{subfigure}%
    ~ 
    \begin{subfigure}{0.5\textwidth}
        \centering
        \includegraphics[width=0.93\textwidth]{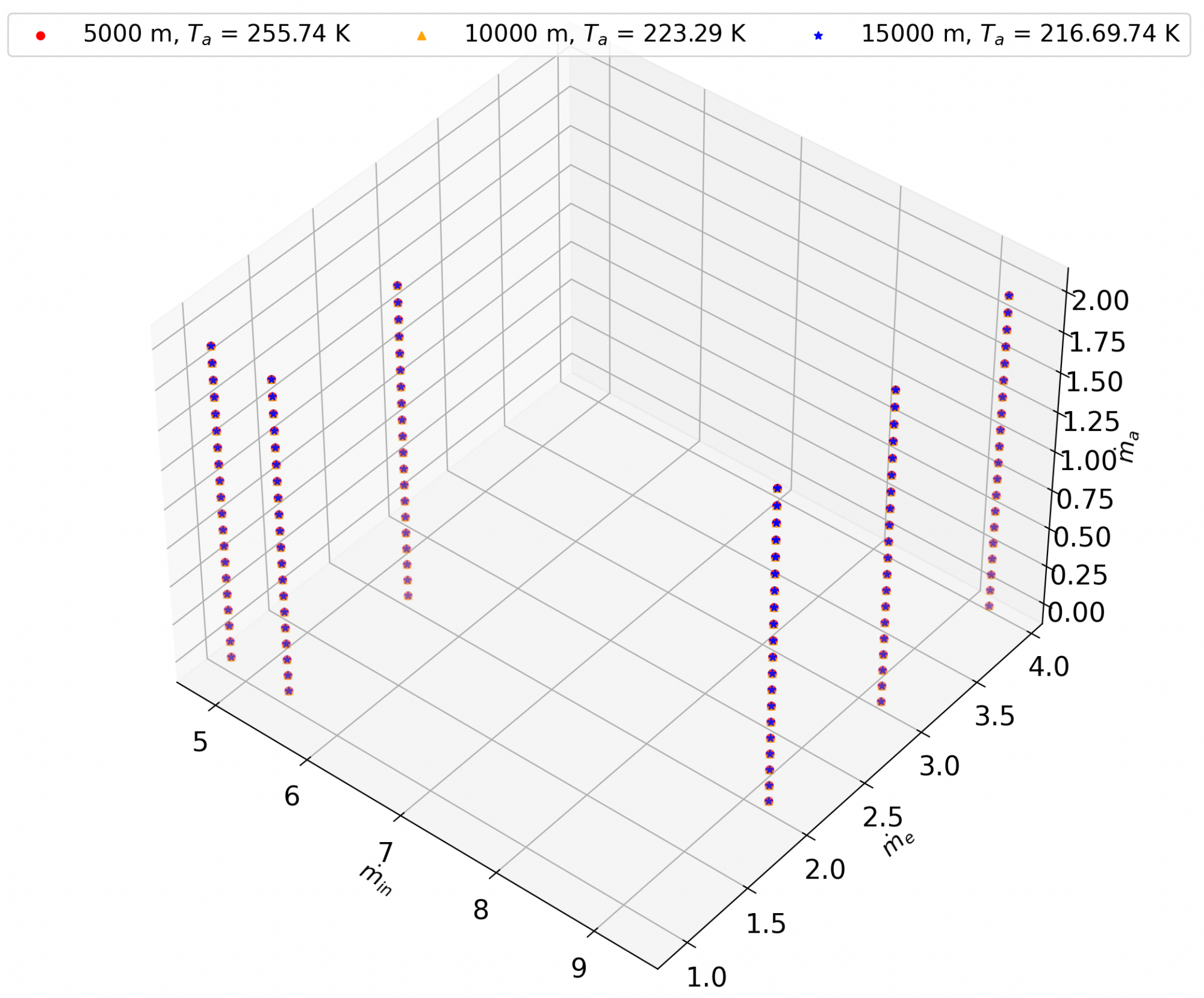}
        \caption{Valid fuel and air flow values for a system operating at different altitudes with a controlled heat exchanger.}
    \end{subfigure}
    \caption{Comparisons between the two different heat exchanger: a heat exhanger with a fixed area and air intake (left), and a heat exchanger where the airflow can be controlled to achieve a  desired output temperature (right).}
    \label{fig:comparison_heat_exchangers}
\end{figure}

At this level of abstraction, it is also possible to compare different component options. For example, consider the case where the fixed heat exchanger shown in Figure~\ref{fig:pacti_aircraft_const_mdot} is replaced by its \emph{controlled} version. The controlled heat exchanger is a more complex sub-system that under the assumption that there is a temperature difference between the hot and cold sides of at least $10$ \unit{\kelvin}, i.e., $T_s - T_a \geq 10$, guarantees that $T_{in} - 5 \leq T_{out} \leq T_{in} + 5$. The implementation may require an operable fixture to modulate $\dot{m}_a$, and a fan when the system is operating a very low speed (e.g., taxiing on the ground). The result of the comparison between the two solutions is shown in Figure~\ref{fig:comparison_heat_exchangers}. Each dot is a valid instance for different combinations of design variables $\dot{m}_{in}$ and $\dot{m}_a$, and operational variables $\dot{m}_e$ (corresponding to thrust level) and $T_a$ (corresponding to flight altitude). We can observe that in the case of the fixed heat exchanger, there is not a combination of design variable values that can satisfy the entire range of the operational variables. In the case of the controlled heat exchanger, instead, it seems sufficient to operate the electric pump in two regimes (low, and high) to cover the entire flight envelope.

Once the exploration is completed, the resulting bounds can be used to select promising instances that can be further optimized. The optimization criteria, or cost function, is related to the quality of the components in the system which, for the generic component $x$, is modeled by the tolerance parameters $\epsilon_x$. Lower values of the tolerance parameters correspond to higher cost. We define a cost function which is the sum of two terms: $c(\mathbf{\epsilon})$ is defined as $||\mathbf{1} - \mathbf{\epsilon}||_2$; the other term is defined by a function $V$ of the temperature bounds. Specifically, $V$ has a large value if either bound violates $G_{S}$, otherwise $V = (T_{e,min} - \underline{T_e})^2 + (T_{out,min} - \underline{T_{out}})^2 + (\overline{T_e} - T_{e,max})^2 +  (\overline{T_{out}} - T_{out,max})^2$. This cost function forces to optimizer to prefer valid instances that span the entire allowed range for the temperature bounds, which correspond to finding the weakest guarantee that still refines the specification. 

For instance, in the case of the controlled heat exchanger, we have selected an instance with altitude $15$ \unit{\kilo\metre}, thrust level $20,000$ \unit{\kilogram}, $\dot{m}_{in} = 9.316$ \unit{\kilogram\per\second}, and $\dot{m}_a = 0.429$ \unit{\kilogram\per\second}. We used the Nelder-Mead optimization methods implemented by the SciPy package~\cite{scipy}, and we set the initial value of all tolerances to 0. We also limit the search space to tolerance values between one and ten percent. After 2000 iteration, the solution found by the optimizer is $\epsilon_{ep,w} = 0.01$, $\epsilon_{ep,t} = 0.09998$, $\epsilon_g = 0.01008$, $\epsilon_{l,w} = 0.1$, $\epsilon_{l,h} = 0.09998$, $\epsilon_{hl} = 0.06214$, $\epsilon_s = 0.01$.

%% file: conclusions.tex
The ability to analyze complex systems early on in the design cycle is essential to system success. In the early stages of design, the development of requirements and their allocation to sub-systems involves a collaborative and iterative effort. Tool support would be beneficial to evaluate trade-offs among multiple viewpoints with respect to the system design and its planned operation. The modeling and analysis tools should be formal, should support compositional design and analysis, and should be efficient and explainable.  

Considering this objective, we presented an agile methodology for designing and analyzing such systems across multiple viewpoints based on a compositional, contract-based modeling paradigm using Pacti~\cite{pacti}. Pacti supports the theory of polyhedral constraints for specifying assume-guarantee contracts. Since this formalism involves linear constraints for specifying assumptions and guarantee constraints, the modeling paradigm is accessible to most stakeholders and appropriate for communicating across stakeholders with diverse expertise and viewpoints the intricacies of space mission design, requirements, and operations formulated in this manner. 

We demonstrated the scalability of several of Pacti's API operations for composing and merging contracts and for computing bounds for variables and linear optimization criteria. The case studies we presented confirm the viability of the general approach. We collected a list of lessons learned from this experiment. Mainly, Pacti is effective at combining contracts, whether these are for system or component specification purposes or for operational requirement purposes. Despite polyhedral algebra being mathematically simple, complex polyhedral contracts can be difficult to understand. In these cases, we found that computing minimum and maximum bounds for contract variables yields valuable information for elaborating operational requirements. In engineering, behavior is typically thought in terms of simulating the state variables as a function of time. Pacti requires systems engineers to reconsider their components as the sets of possible behaviors they can reflect;
on the other hand, Pacti's contract algebra provides powerful tools to help understand bounded behavior. For example, developing a time-based simulation model involves a risk that one could get lost in the details and lose track of the overall modeling objective. In contrast, Pacti's polyhedral contract algebra coerces the system engineer to think about which aspects of behavior are important to characterize with bounds.

\section*{Acknowledgments}

JPL/Caltech Copyright: © 2023 California Institute of Technology. Government sponsorship acknowledged. The research was carried out at the Jet Propulsion Laboratory, California Institute of Technology, under a contract with the National Aeronautics and Space Administration (80NM0018D0004). This work was partially supported by the JPL Researcher On Campus (JROC) program for which we gratefully acknowledge Prof. Richard M. Murray's support. This work was also partially supported by NSF and ASEE through an eFellows postdoctoral fellowship.